\documentclass{aa}  

\usepackage{natbib}
\usepackage{graphicx}
\usepackage{txfonts}
\usepackage[draft]{hyperref}
\usepackage{color}
\usepackage{subfig}

\begin{document} 
  \title{Orbital and spectral analysis of the benchmark brown dwarf HD\,4747B\thanks{Based on observations made with the instrument SPHERE (Prog. ID 198.C-0209) and NaCo (Prog. ID 081.C-0917(A)) at the Paranal observatory and with the CORALIE echelle spectrograph mounted on the 1.2 m Swiss telescope at La Silla Observatory.}}
  \author{S. Peretti\inst{\ref{Geneva}}
  \and D. S{\'e}gransan\inst{\ref{Geneva}}
  \and B. Lavie \inst{\ref{Geneva}}
  \and S. Desidera \inst{\ref{Padova}}
  \and A.-L. Maire\inst{\ref{Max_Planck}}
  \and V. D'Orazi \inst{\ref{Padova}}
  \and A. Vigan \inst{\ref{Marseille}}
  \and J.-L. Baudino \inst{\ref{LESIA},\ref{Oxford}}
  \and A. Cheetham \inst{\ref{Geneva}}
  \and M. Janson \inst{\ref{Max_Planck},\ref{Stockholm}}
  \and G. Chauvin \inst{\ref{Grenoble}}
  \and J. Hagelberg \inst{\ref{Grenoble}}
  \and F. Menard \inst{\ref{Grenoble}}
  \and K. Heng \inst{\ref{Bern}}
  \and S. Udry \inst{\ref{Geneva}}
  \and A. Boccaletti \inst{\ref{LESIA}}
  \and S. Daemgen \inst{\ref{ETH}}
  \and H. Le Coroller \inst{\ref{Marseille}}
  \and D. Mesa \inst{\ref{Padova}}
  \and D. Rouan \inst{\ref{LESIA}}
  \and M. Samland \inst{\ref{Max_Planck}}
  \and T. Schmidt \inst{\ref{LESIA}}
  \and A. Zurlo \inst{\ref{Marseille},\ref{Santiago1},\ref{Santiago2}}
  \and M. Bonnefoy \inst{\ref{Grenoble}}
  \and M. Feldt \inst{\ref{Max_Planck}}
  \and R. Gratton \inst{\ref{Padova}}
  \and A-M. Lagrange \inst{\ref{Grenoble}}
  \and M. Langlois \inst{\ref{Marseille},\ref{CNRS_Lyon}}
  \and M. Meyer \inst{\ref{ETH},\ref{Michigan}}
  \and M. Carbillet \inst{\ref{CNRS_Lagrange}}
  \and M. Carle \inst{\ref{Marseille}}
  \and V. De Caprio \inst{\ref{Napoli}}
  \and L. Gluck \inst{\ref{Grenoble}}
  \and E. Hugot \inst{\ref{Marseille}}
  \and Y. Magnard \inst{\ref{Grenoble}}
  \and T. Moulin \inst{\ref{Grenoble}}
  \and A. Pavlov \inst{\ref{Max_Planck}}
  \and J. Pragt \inst{\ref{NOVA}}
  \and P. Rabou \inst{\ref{Grenoble}}
  \and J. Ramos \inst{\ref{Max_Planck}}
  \and G. Rousset \inst{\ref{LESIA}}
  \and A. Sevin \inst{\ref{LESIA}}
  \and C. Soenke \inst{\ref{ESO}}
  \and E. Stadler \inst{\ref{Grenoble}}
  \and L. Weber \inst{\ref{Geneva}}
  \and F. Wildi \inst{\ref{Geneva}}
          }

  \institute{Observatoire astronomique de l'Universit{\'e} de Gen{\`e}ve, 51 Ch des Maillettes, 1290 Sauverny, Switzerland  \label{Geneva} \\\email{sebastien.peretti@unige.ch}
  \and INAF - Osservatorio Astronomico di Padova, Vicolo dell'Osservatorio 5, 35122, Padova, Italy \label{Padova}
  \and Max Planck Institute for Astronomy, Koenigstuhl 17 69117 Heidelberg Germany \label{Max_Planck}
  \and Laboratoire d'Astrophysique de Marseille, 38, rue Fr{\'e}d{\'e}ric Joliot-Curie 13388 Marseille cedex 13 FRANCE \label{Marseille}
  \and LESIA,  Observatoire  de  Paris,  PSL  Research  University,  CNRS, Sorbonne  Universit{\'e}s,  UPMC  Univ.  Paris  06,  Univ.  Paris  Diderot, Sorbonne Paris Cit{\'e}, 5 place Jules Janssen, 92195 Meudon, France \label{LESIA}
  \and Department of Physics, University of Oxford, Oxford, UK \label{Oxford}
  \and Department of Astronomy, Stockholm University, AlbaNova University Center, 106 91 Stockholm, Sweden \label{Stockholm}
  \and Univ. Grenoble Alpes, CNRS, IPAG, 38000 Grenoble, France \label{Grenoble}
  \and University of Bern, Center for Space and Habitability, Sidlerstrasse 5, CH-3012, Bern, Switzerland \label{Bern}
  \and Universit{\'e} Cote d'Azur, OCA, CNRS, Lagrange, France \label{CNRS_Lagrange}
  \and Institute for Astronomy, ETH Zurich, Wolfgang-Pauli-Strasse 27, 8093 Zurich, Switzerland \label{ETH}
  \and INAF - Osservatorio Astronomico di Capodimonte, Salita Moiariello 16, 80131 Napoli, Italy \label{Napoli}
  \and Centre de Recherche Astrophysique de Lyon, CNRS, Universit{\'e} Lyon 1, 9 avenue Charles Andr{\'e}, F-69561 Saint Genis-Laval-Cedex, France \label{CNRS_Lyon}
  \and The University of Michigan, Ann Arbor, MI 48109, USA \label{Michigan}
  \and NOVA Optical Infrared Instrumentation Group, Oude Hoogeveensedijk 4, 7991 PD Dwingeloo, The Netherlands \label{NOVA}
  \and European Southern Observatory (ESO), Karl-Schwarzschild-Str. 2, 85748 Garching, Germany \label{ESO}
  \and N{\'u}cleo de Astonom{\'i}a, Facultad de Ingenier{\'i}a, Universidad Diego Portales, A. Ejercito 441, Santiago, Chile \label{Santiago1}
  \and Escuela de Ingenier\'ia Industrial, Facultad de Ingenier\'ia y Ciencias, Universidad Diego Portales, Av. Ejercito 441, Santiago, Chile \label{Santiago2}
  }

\date{}

 \abstract
{The study of high contrast imaged brown dwarfs and exoplanets depends strongly on evolutionary models. To estimate the mass of a directly imaged substellar object, its extracted photometry or spectrum is used and adjusted with model spectra together with the estimated age of the system. These models still need to be properly tested and constrained. HD 4747B is a brown dwarf close to the H burning mass limit, orbiting a nearby ($d=19.25\pm0.58pc$), solar-type star (G9V) and has been observed with the radial velocity method over almost two decades now. Its companion was also recently detected by direct imaging, allowing a complete study of this particular object.}
{We aim to fully characterize HD 4747B by combining a well constrained dynamical mass and a study of its observed spectral features in order to test evolutionary models for substellar objects and characterize its atmosphere.}
{We combine the radial velocity measurements of HIRES and CORALIE taken over two decades and high contrast imaging of several epochs from NACO, NIRC2 and SPHERE to obtain a dynamical mass. From the SPHERE data we obtain a low resolution spectrum of the companion from Y to H band, as well as two narrow band-width photometric measurements in the K band. A study of the primary star allows in addition to constrain the age of the system as well as its distance.}
{Thanks to the new SPHERE epoch and NACO archival data combined with previous imaging data and high precision radial velocity measurements, we have been able to derive a well constrained orbit. The high eccentricity ($e=0.7362\pm0.0025$) of HD 4747B is confirmed, and the inclination as well as the semi-major axis are derived ($i=47.3\pm1.6^\circ$, $a=10.01\pm0.21$ au). We derive a dynamical mass of $m_{\rm B}=70.0\pm1.6$ $\rm M_{Jup}$ which is higher than a previous study, but in better agreement with the models. By comparing the object with known brown dwarfs spectra, we derive a spectral type of L9 and an effective temperature of $1350\pm50$ K. With a retrieval analysis we constrain the oxygen and carbon abundances and compare them with the ones from the HR 8799 planets.}
{}

 \keywords{stars: binaries: general -- stars: binaries: spectroscopic -- stars: binaries: visual -- stars: low-mass, brown dwarfs -- techniques: radial velocities -- techniques: adaptive optics}
\maketitle
\section{Introduction}
Brown dwarfs (BD) are substellar objects, not massive enough to sustain hydrogen burning, with masses below $\sim$75 Jupiter masses \citep[e.g.][]{Burrows1997}. Since the first brown dwarfs were discovered in 1995 \citep{Nakajima1995,Oppenheimer1995,Rebolo1995} by imaging, only few of these objects were detected around sun-like stars with respect to the number of planetary mass objects, and binaries \citep[e.g.][]{Anderson2011,Sahlmann2011b,Siverd2012,Bayliss2017}. Numerous studies based on transiting, radial velocity and astrometry methods have indeed demonstrated that the "brown dwarf desert" is observed at separations below 10 au and that brown dwarfs have a frequency around sun like stars of less than $1\%$ \citep{Halbwachs2000,Marcy2000,Grether2006,Sahlmann2011,Wilson2016}. However, a recent study showed that this "desert" might exists only for separations smaller than $\sim0.1$ - $\sim0.2$ au \citep[e.g.][]{Troup2016}. At larger separation, BD are found as abundantly as very low mass stars. However the much wider range in spectral types in comparison with RV's surveys probed by \citet{Troup2016} might explain the high number of BD found.

Due to observing biases, direct imaging is much better suited to exploring the outer reaches of stellar systems to search for such brown dwarf companions. A few brown dwarfs have been detected by direct imaging \citep[e.g.][]{Thalmann2009,Biller2010,Chauvin2010}, allowing to constrain their effective temperature and atmospheric properties thanks to spectrophotometry analysis \citep[e.g.][]{Maire2016b,Vigan2016} and the age of the system through the host stars. To determine the mass of an imaged BD companion, the key parameter for the evolution of substellar objects, we usually rely on evolutionary models \citep[e.g.][]{Baraffe2003}. These models still need to be tested and properly calibrated through observations. To achieve this, observations of objects for which we can constrain independently the age, the effective temperature and the mass are needed.

Free floating BD have been detected and provided high resolution spectra, allowing to better understand these objects \citep[see e.g.][and ref. therein]{Kirkpatrick2005,Helling2014}, but independent mass and age estimations cannot be derived in most cases. BD companions to solar type stars have been discovered at wide orbits \citep[e.g.][and ref. therein]{Burgasser2007b,Pinfield2012,Burningham2013}, allowing an age determination from their host stars and effective temperature from their spectra. However their wide orbits prevent from a dynamical mass measurement. Dynamical masses of BD have been determined for BD pairs \citep[e.g.][]{DupuyLiu2017}, but no independent age estimation could be extracted for these sub-stellar systems. For some objects we have high resolution spectra and dynamical masses, but these often also lack independent age measurements \citep[e.g.][]{King2010,Line2015}. The majority of directly imaged brown dwarf companions from high contrast imaging surveys have been detected around young and massive stars, as they are brighter at young ages \citep{Chabrier2000}. These detections can provide age estimations of the system from the stellar host, as well as astrometric orbits, but model-independent dynamical masses cannot be easily obtained due to the difficulties in achieving precise radial velocities of young, massive stars \citep[e.g.][]{Galland2005}. In addition the orbital period of these directly imaged brown dwarfs are usually of decades if not centuries, and therefore obtaining a complete orbit will need long term monitoring.

HD 4747B is a perfect candidate to test the evolutionary  models. It is orbiting a late G type star (HD 4747), which was observed through two decades by the HIRES instrument at the Keck \citep{Vogt1994}, as well as with the CORALIE spectrograph \citep{Queloz2000} to obtain radial velocities. Two thirds of the orbit has been already completed, and the important periastron passage was covered. This allowed a minimum mass determination of the companion that pointed towards a brown dwarf at large separation \citep{Nidever2002,Sahlmann2011}. Thanks to the high precision of the RV data and the number of points, the minimum mass is well constrained and only a few astrometric points can provide a high precision orbit and dynamical mass. HD 4747B was directly imaged for the first time by \citet{Crepp2016} who confirmed its substellar nature and gave a dynamical mass estimation of $m_{\rm B}=60.2\pm3.3$ $\rm M_{Jup}$\footnote{During the referee process of this paper, additional results from observations taken with GPI were published by \citet{Crepp2018}. The dynamical mass has been updated at $m_{\rm B}=65.3^{+4.4}_{-3.3}$ $\rm M_{Jup}$.}. However the isochronal mass estimate of $m_{\rm B}=72^{+3}_{-13}$ $\rm M_{Jup}$ was in marginal agreement with its dynamical mass. A color-magnitude diagram led to a late L spectral type.

To improve the orbital parameters and characterize the atmosphere of HD 4747B, we have observed HD 4747 with the SPHERE instrument installed on the VLT \citep[Spectro-Polarimetric High-contrast Exoplanet REsearch;][]{Beuzit2008} in December 2016. In Sect. \ref{star} we detail our analysis of the host star, which gives the age and distance of the system. We describe in Sect. \ref{observations} the observations we used in this paper. In Sect. \ref{imaging_data} we report on the extraction of spectrophotometry and astrometry from the SPHERE images and we describe also the analysis of an archival NACO dataset in which we could detect the companion. Sect. \ref{orbit} presents our orbital analysis combining radial velocity and imaging data. Sect. \ref{spec_phot} describes the spectrophotometric analysis, by comparing our extracted spectra with real objects, models and a retrieval analysis, and we conclude in Sect. \ref{conclusion}.

\section{Host star properties}\label{star}
As a close ($d=19.3\pm0.6$ pc, this paper) solar type star, HD 4747 was extensively studied in the literature. Several spectroscopic analyzes were performed by different groups (Table \ref{t:spec}), indicating an effective temperature around 5300-5400 K, a gravity log$g$=4.5-4.65 and a mildly subsolar metallicity ([Fe/H] about -0.2). Similar values were also derived from Str\"omgren photometry.

\begin{table*}[!t]
\centering
\caption{Spectroscopic parameters of HD 4747}\label{t:spec}
\begin{tabular}{llllll}
\hline\hline
$T_{\rm eff}$ [K]  & $\log$ g  [dex] & $\xi$ [km s$^{-1}$]& [Fe/H] [dex]& v $\sin$ i [km s$^{-1}$]& Reference \\
\hline
5337$\pm$80 &  4.58$\pm$0.10  &  0.85$\pm$0.20  & $-$0.25$\pm$0.07 & 2.3$\pm$1.0 & \cite{fuhrmann2017} \\
5347        &                 &                 & $-$0.21          &             & \cite{mortier2013} \\                  
5422$\pm$75 &  4.61           &                 & $-$0.15          &             & \cite{casagrande2011} \\
5305        &  4.56           &                 & $-$0.24          & 2.1         & \cite{brewer2016} \\
5335        &  4.65           &                 & $-$0.22          & 1.1         & \cite{valenti2005} \\
5316$\pm$38 &  4.48$\pm$0.10  &  0.79$\pm$0.06  & $-$0.21$\pm$0.05 & 0.79        & \cite{santos2005} \\
5340$\pm$40 &  4.65$\pm$0.06  &                 & $-$0.22$\pm$0.04 & 1.1$\pm$0.5 & \cite{Crepp2016} \\
\hline
5400$\pm$60    &  4.60$\pm$0.15  &  0.75$\pm$0.20           & $-$0.23$\pm$0.05          & 2.0$\pm$1.0           & this paper \\
\hline\hline
\end{tabular}
\end{table*}

We used an archival spectrum of HD 4747 taken with the visible high resolution spectrograph FEROS\footnote{PI: Rolf Chini, Prog. ID 095.A-9029} \citep{Kaufer1997}, to re-derive its stellar parameters. The effective temperature ($T_{\rm eff}$), the gravity (log$g$), the microturbulence ($\xi$) and the metallicity ([Fe/H]) were retrieved using the standard approach described in \citet{dorazi2017}. We obtain $T_{\rm eff}$=5400$\pm$60 K, log$g$=4.60$\pm$0.15 dex, $\xi$=0.75$\pm$0.2 km s$^{-1}$ and [Fe/H]=$-$0.23$\pm$0.05 dex. Our results, also reported in Table \ref{t:spec}, are fully consistent with those from the literature. Moreover, we have derived abundances for elements produced in the slow ($s$) neutron-capture process, namely yttrium, barium and lanthanum. As done in all our previous investigations, we have carried out spectral synthesis calculations, including isotopic splitting and hyperfine structure as needed (see e.g., \citealt{dorazi12}; \citealt{dorazi2017}). We have detected a modest enhancement in $s$-process element abundances, finding  [Y/Fe]=+0.30$\pm$0.15, [Ba/Fe]=+0.35$\pm$0.20 dex and [La/Fe]=+0.20$\pm$0.12 dex. This might be in principle the signature of a weak contamination from a companion during the asymptotic giant branch (AGB) phase but no indication of the presence of a white dwarf companion at any separation was detected and the HD 4747B spectrum is not compatible with a white dwarf spectrum. Most importantly, we have also derived carbon abundance (pollution from low mass AGB stars results in enhanced C abundances), by synthesizing the CH band at 4300 $\AA$: we have obtained a solar scaled abundance of [C/Fe]=0$\pm$0.13 dex, which points against the AGB pollution scenario. Moreover, when observation uncertainties are taken into account, the $s$ process element abundances for HD 4747 are consistent with the scattered distribution as revealed from field stars (see e.g., \citealt{Bensby2014}).

A crucial information for the characterization of the low-mass companion HD 4747B is the stellar age. Literature shows quite discrepant values, due to different dating techniques employed. Indeed, as expected for a late G/early K dwarf close to main sequence, isochrone fitting allows only a poor constraint on stellar age (basically upper limits of about 7-9 Gyr). The spectrum shows no lithium, putting a lower limit to stellar age at about 700 Myr.

\begin{figure}[!h]
\includegraphics[width=0.97\columnwidth]{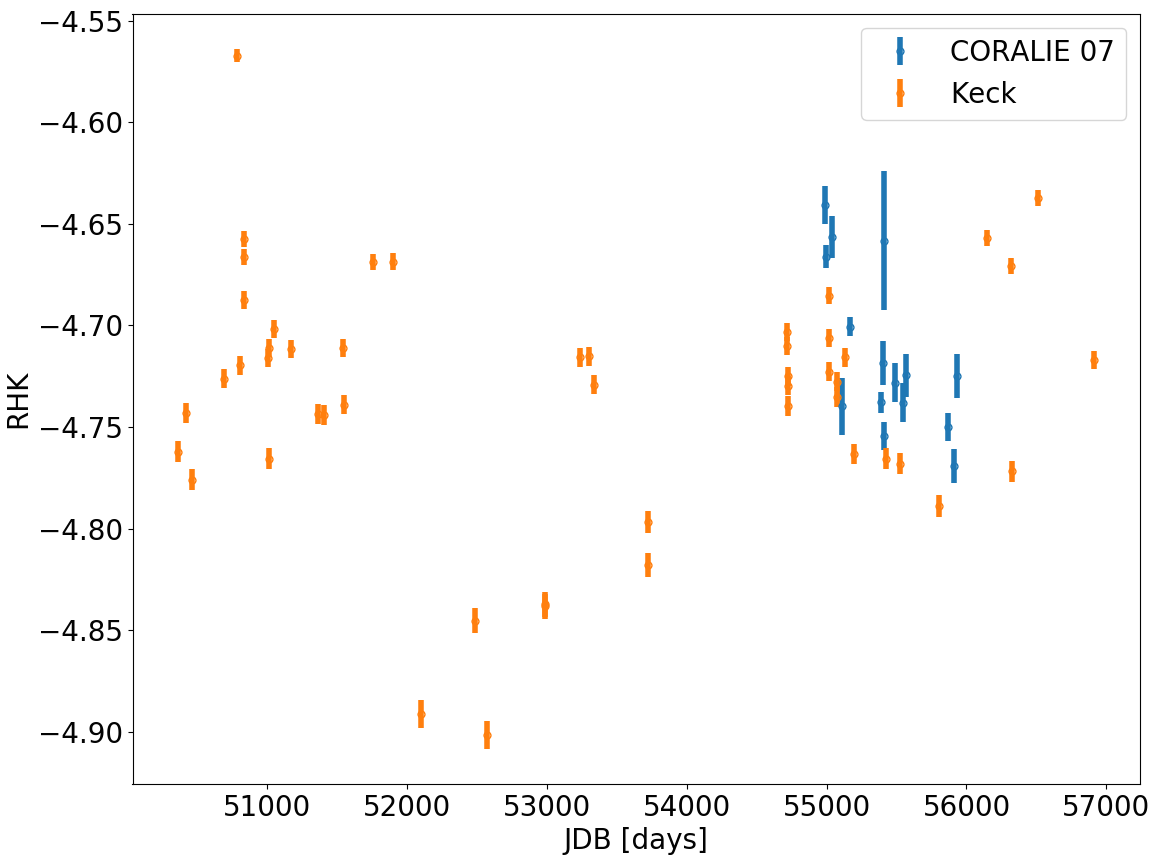}
\caption{$\log(R_{\rm HK})$ from the Keck and CORALIE data}
\label{rhk}
\end{figure}
The best constraints are derived from coronal and chromospheric emission. Using the Keck S index values as reported in \citet{butler2017} and B-V from Hipparcos, we derived a median values of $\log R_{\rm HK}=-4.725$ with a r.m.s of 0.023 dex. From the CORALIE dataset described in Sect. \ref{RV_sect}, $\log R_{\rm HK}=-4.718$ is obtained, with r.m.s. of 0.007 dex (see Fig. \ref{rhk}).

These values are intermediate between the activity levels of the Hyades and M67 open clusters and clearly above that of the Sun. The extension of the Keck dataset (50 epochs over almost 18 yr) ensures that intrinsic variability is averaged out. Using the age-$\log R_{\rm HK}$ calibration by \cite{mamajek2008} an age of 2.3 Gyr is obtained. 
A very similar age, 2.1 Gyr, is obtained from X-ray emission \citep{kasova2011}, when applying \cite{mamajek2008} calibration. On the basis of these results we favour an age of 2.3 Gyr with upper and lower limits of 0.9-3.7 Gyr as the activity level of HD 4747 is clearly distinct from the Hyades members one at the young side and from the Sun's one even at its maximum of the activity cycle at the old side. We also used the \cite{mamajek2008} calibration in order to compute the expected rotational period of HD 4747 and found $P_{\rm rot}=25.8\pm4.1$ days. 

\begin{figure}[!htb]
\centering
	\subfloat[Periodogram of $\log R_{\rm HK}$.]{\includegraphics[width=0.97\columnwidth]{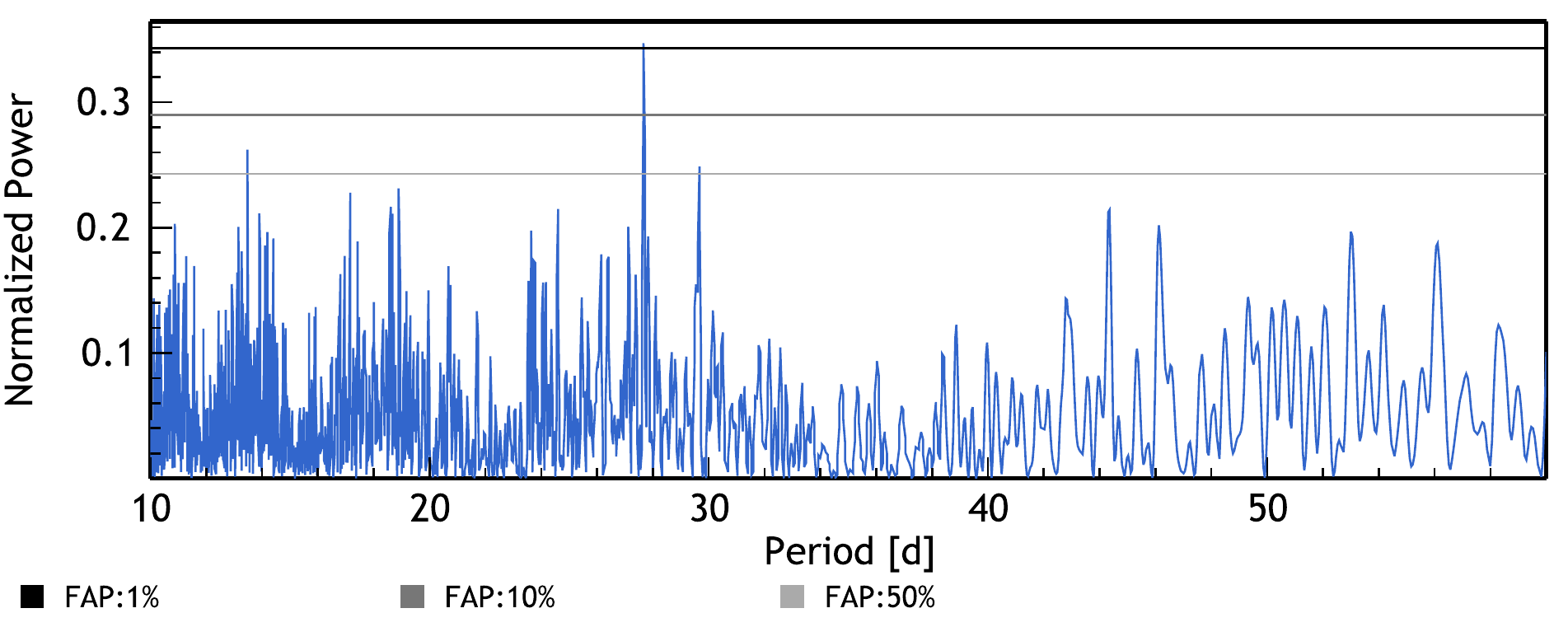}\label{periodogram_rhk}}
\hspace{0.02 cm}
	\subfloat[Periodogram of the residual of the RV's data.]{\includegraphics[width=0.97\columnwidth]{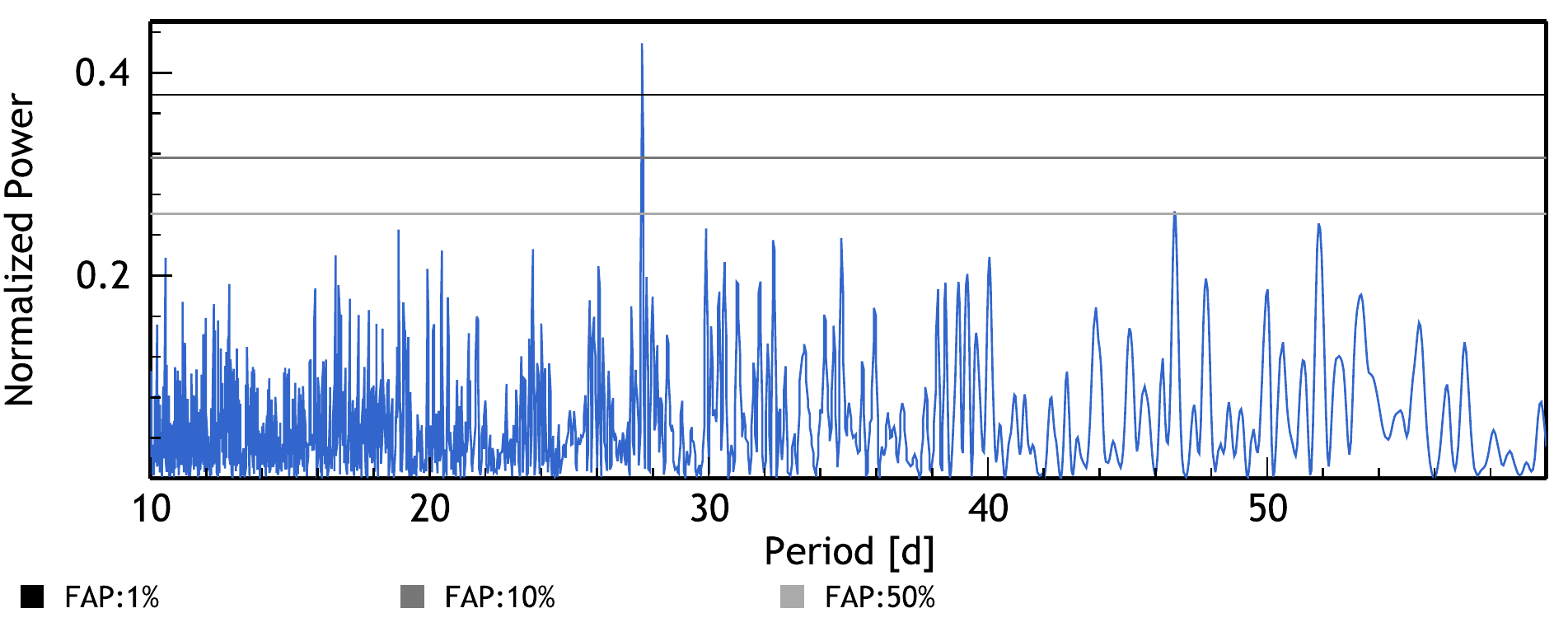}\label{periodogram_vrad}}
\caption{GLS periodogram \citep{ZK09} of the $\log R_{\rm HK}$ and the residu of the radial velocity data from the HIRES and CORALIE spectrographs (see Sect. \ref{orbit} for the orbital parameters)\protect\footnotemark.}\label{periodograms}
\end{figure}

By looking at the periodograms of the $\log R_{\rm HK}$ (Fig. \ref{periodogram_rhk}) we find a clear signal at 27.7 days. A signal is observed at the same period in the periodogram of the residual of the radial velocity data (Fig. \ref{periodogram_vrad}). We favor then a rotational period of $P_{\rm rot}=27.7\pm0.5$ days.

\footnotetext{The graphs have been done by using a set of online tools hosted by the Data \& Analysis Center for Exoplanets (DACE), which is available at: \url{http://dace.unige.ch}}

\cite{fuhrmann2017} noted that the kinematic parameters support membership in the Hyades stream but also the much lower metallicity of HD 4747 with respect to the Hyades.
They also mention a possible inconsistency between the low metallicity and the activity level (moderately young age) and speculated about the possibility of accretion of angular momentum to make the star appearing younger \citep[see ][for a description of this mechanism in the case of GJ504]{dorazi2017}. Conclusive proof that this mechanism is really at work in the case of HD 4747 is much challenging to obtain with respect to GJ504, because of the different main sequence lifetimes. However, to further shed light on this possibility, we investigated whether the age/metallicity obtained for HD 4747 is really peculiar using the extensive database of the Geneva-Copenhagen survey \citep{nordstrom2004}. Exploiting the latest age and metallicity determination by \citet{casagrande2011}, we searched for stars with ages within our adopted upper limit for HD 4747 and metallicity within $\pm$0.05 dex. Hundreds of objects were returned, typically F type stars, indicating that the age/metallicity combination of HD 4747 is not particularly unusual. Extended moving groups like the Hyades stream were also shown to host a mixture of stellar populations \citep{famaey2005}, then HD 4747 is not particularly anomalous also from this point of view. We then conclude that, while accretion events altering the angular momentum evolution and then the age from activity cannot be firmly ruled out, there are no specific indications supporting that this kind of evolution affected our target. Therefore, we dismiss such an hypothesis and we adopt in the following the age from coronal and chromospheric activity.

\begin{table}
\begin{small}
\caption{Stellar parameters of HD 4747}\label{t:param}
\begin{tabular}{lcl}
\hline\hline
Parameter      & Value  & Ref \\
\hline
V (mag)                   &    7.15   & Hipparcos \\
B$-$V (mag)               &   0.769$\pm$0.009   & Hipparcos \\
V$-$I (mag)               &    0.82$\pm$0.02         & Hipparcos \\
J (mag)                   &     5.813$\pm$0.021  & 2MASS \\
H (mag)                   &     5.433$\pm$0.049  & 2MASS \\
K (mag)                   &     5.305$\pm$0.029  & 2MASS \\
Parallax (mas)            &    53.51$\pm$0.53 & \citet{van-Leeuwen2007} \\
Parallax (mas)            &    50.37$\pm$0.55 & \citet{gaia_dr1} \\  
Parallax (mas)            &    51.94$\pm$1.57 & Adopted for this paper \\
$\mu_{\alpha}$ (mas\,yr$^{-1}$)  & 516.92$\pm$0.55  & \citet{van-Leeuwen2007} \\
$\mu_{\delta}$ (mas\,yr$^{-1}$)  & 120.05$\pm$0.45  & \citet{van-Leeuwen2007} \\
$\mu_{\alpha}$ (mas\,yr$^{-1}$)  & 515.509$\pm$0.032  & \citet{gaia_dr1} \\
$\mu_{\delta}$ (mas\,yr$^{-1}$)  & 125.472$\pm$0.031  & \citet{gaia_dr1} \\
$T_{\rm eff}$ (K)      &  5400$\pm$60 & this paper \\
$\log g$             &    4.60$\pm$0.15 & this paper \\
${\rm [Fe/H]}$       &   $-$0.23$\pm$0.05   & this paper \\
$v \sin i $  (km\,s$^{-1}$)          & 2$\pm$1 & this paper \\
$\log R_{\rm HK}$         &   -4.725    & Keck \\
$\log R_{\rm HK}$         &   -4.718    & CORALIE \\
$P_{\rm rot}$ (days)   &  27.7$\pm$0.5  & this paper \\
$\log L_{\rm X}/L_{\rm bol}$   & -5.52      & \citet{kasova2011} \\
EW Li (m\AA)         &   0    & this paper \\
Age (Gyr)            &    2.3$\pm$1.4 & this paper  \\
$M_{\rm star}$ ($\rm M_{\odot}$)   &  0.856$\pm$0.014  & this paper \\ 
$R_{\rm star}$ ($\rm R_{\odot}$)   &  0.769$\pm$0.016  & this paper \\
\hline\hline
\end{tabular}
\end{small}
\end{table}

The stellar mass was derived from isochrones using the PARAM web interface\footnote{http://stev.oapd.inaf.it/cgi-bin/param\_1.3} \citep{param} isolating the age range allowed by indirect methods \citep[see ][]{desidera2015}. Our spectroscopic $T_{\rm eff}$ and [Fe/H] were adopted. The outcome is slightly dependent on the adopted distance.
Hipparcos and Gaia DR1 distance are formally discrepant at more than 4 $\sigma$ level, possibly due to the unaccounted orbital motion due to the brown dwarf companion\footnote{This is actually supported also by the formally significant differences in proper motion in various catalogs}. Adopting the average of the two measurements, the stellar mass results in 0.856$\pm$0.014 $\rm M_{\odot}$. 

Finally, the star was observed with Spitzer and Herschel, resulting in no detectable IR excess \citep{gaspar2013}. The stellar parameters are summarized in Table \ref{t:param}.

\begin{figure}[!h]
\includegraphics[width=0.97\columnwidth]{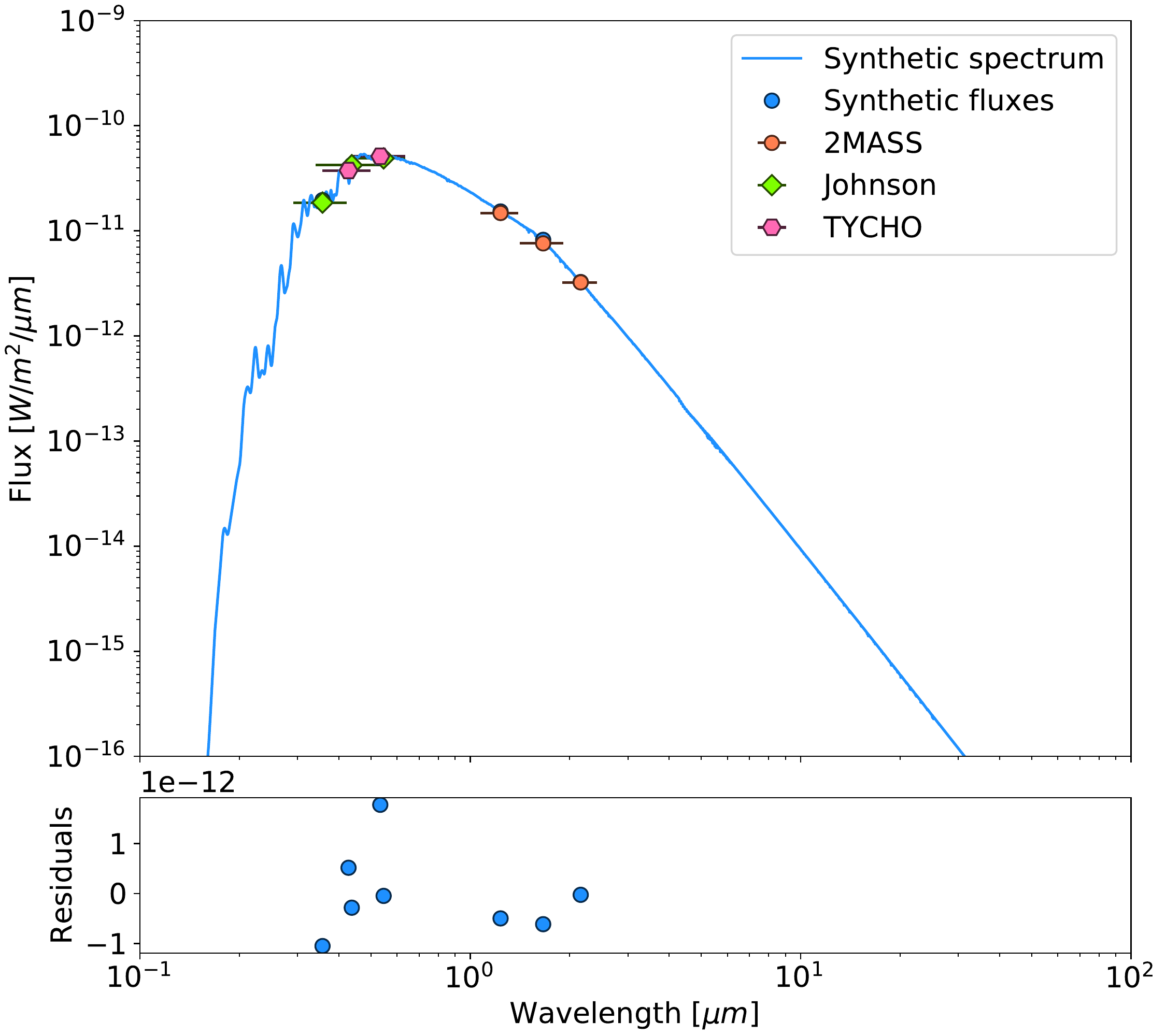}
\caption{BT-NextGen synthetic spectrum of HD 4747, scaled to match SED of optical and mid-infrared photometry}
\label{sed}
\end{figure}

\section{Observations}\label{observations}
Radial velocity and direct imaging observations were combined to constrain the orbit of HD 4747B. The good orbital coverage of the radial velocity time series allows to constrain HD 4747B's period, minimum mass, eccentricity,  argument of periastron $\omega$ and  time of periastron passage $T_{0}$. Combined to the few direct imaging observations spread over $\sim27$\% of the period we are able to retrieve both  longitude of ascending node $\Omega$ and the orbit inclination $i$. In addition, the SPHERE direct imaging observations allow us to obtain a spectrum of the brown dwarf companion.

\begin{figure*}[!t!!h]
\centering
	\subfloat[IRDIS reduction using median combine only]{\includegraphics[width=0.485\columnwidth]{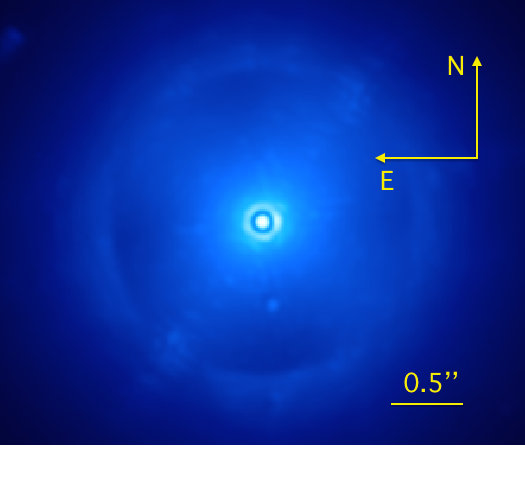}\label{median_combine}}
\hspace{0.02 cm}
	\subfloat[IRDIS reduction using radial-profile subtraction]{\includegraphics[width=0.485\columnwidth]{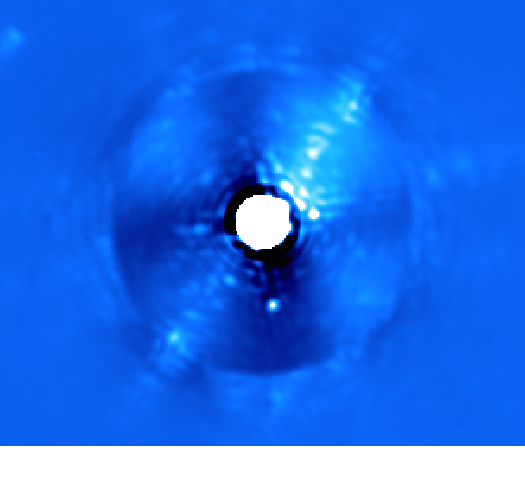}\label{radial_profile}}
\hspace{0.02 cm}
	\subfloat[IRDIS reduction using ADI (TLOCI)]{\includegraphics[width=0.485\columnwidth]{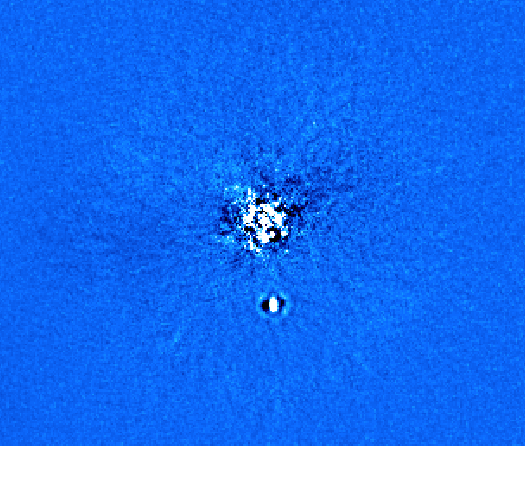}\label{tloci}}
\hspace{0.02 cm}
	\subfloat[IRDIS reduction using SDI and simple derotation]{\includegraphics[width=0.485\columnwidth]{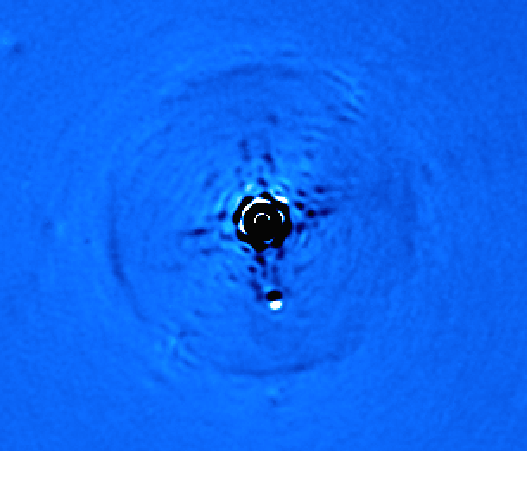}\label{sdi}}
\caption{IRDIS different reductions of HD 4747. The companion is observed in each reduction. The north and east directions and the scale are indicated on the image (a).}\label{IRDIS_reduction}
\end{figure*}

\subsection{Radial velocity}\label{RV_sect}
HD 4747 has been observed since 1999 with the CORALIE spectrograph \citep{Queloz2000} installed on the 1.2 m EULER Swiss telescope at La Silla observatory (Chile). Since its installation in 1998, CORALIE has undergone two upgrades ( in 2007 and 2014) that  introduced small RV offsets that vary from star to star. In the RV modeling procedure, we adjust the RV offsets corresponding to C98 for the data prior to 2007, to C07 for the period between 2007-2014, and C14 for the data acquired since 2014. 

We combined these data with observations conducted between 1996 and 2014 \citep{butler2017}, with the High Resolution Echelle Spectrometer (HIRES) at Keck \citep{Vogt1994}. All of the observations have been taken with high signal to noise ratios, thanks to the relative brightness of the primary star (V=7.155). The high eccentricity ($e=0.736\pm0.002$) of the companion and the fact that we have the periastron passage with the HIRES observations, together with the very long time spanned (20 years) and the high precision of the data allow us to constrain very well the orbital elements, even if the orbit is not complete ($57\%$ of the complete orbit is covered).

\subsection{Direct imaging} \label{direct_imaging}
HD 4747 was observed on the $12^{th}$ of December 2016 and the $28^{th}$ of September 2017 as part of SHINE (SpHere INfrared survey for Exoplanets). SPHERE \citep[Spectro-Polarimetric High-contrast Exoplanet REsearch;][]{Beuzit2008}, is an extreme adaptive optics system \citep{Fusco2014} installed on the VLT in Paranal (Chile). We used SPHERE in its IRDIFS-EXT mode which consists in two instruments working simultaneously, a dual-band imager and spectrograph \citep[IRDIS;][]{Dohlen2008}, and an integral field spectrograph \citep[IFS;][]{Claudi2008}. We used IRDIS in the dual band imaging mode \citep[DBI;][]{Vigan2010} in K12 ($\lambda_{\rm K1}^{\rm c}=2.103$ $\mu$m, FWHM$_{\rm K1}=0.102$ $\mu$m; $\lambda_{\rm K2}^{\rm c}=2.255$ $\mu$m, FWHM$_{\rm K2}=0.109$ $\mu$m), and IFS in YH band ($0.95 - 1.65$ $\mu$m, average spectral resolution per FWHM, R=29). As HD 4747 has a declination very close to the Paranal's latitude, the star passes almost at the zenith. This results in a small field rotation in our observation ($7.3^\circ$), but thanks to the moderate contrast of the companion in the infrared ($\Delta K_1=9.11\pm0.15$), we managed to extract the spectrophotometry and astrometry from our data with a very high accuracy (see Sect. \ref{spec_phot}).

The observations were done using the standard SHINE strategy. The data sequences consist then of the science coronagraphic observations in IRDIFS-EXT mode with DIT=32s, followed by a star centering dataset with same DITs and with the satellite spots, induced by the deformable mirror, activated in order to recover the position of the star behind the coronagraph \citep{Langlois2013}. Then a flux calibration dataset was produced with DIT=2 s, with the star offset from the coronagraph, and a neutral density filter (ND\_2.0) in order to prevent saturation. A set of sky frames have also been taken just after the sequence. The rest of the calibrations (flats, darks and spectral calibrations) were done after the end of the night with the instrument internal calibration hardware. The astrometric calibration (True North and pixel scale) was done on sky with the SPHERE GTO standard procedure \citep{Maire2016}. The 2017 epoch was done in the same way as the 2016 but with a shorter coronagraphic sequence as it was mainly aiming to better constrain the orbit.

\subsection{Archival NACO observation} \label{NACO}
An archival NACO dataset from the $7^{th}$ of september 2008 was also reprocessed to search for and possibly retrieve the separation and position angle of the companion. It was taken in the SDI+4 mode \citep{Maire2014} which combines the SDI mode of NACO with a four quadrant phase mask coronagraph \citep{Boccaletti2004}. The data was taken in the "double roll subtraction" procedure\footnote{See NACO manual for Period 81 and 82: \\ \url{https://www.eso.org/sci/facilities/paranal/instruments/naco/doc/VLT-MAN-ESO-14200-2761\_v81-2.pdf}} with $5^\circ$ of rotation offsets every 5 images, in order to subtract the speckles linked to pupil aberrations. We obtained a field rotation of $95^\circ$, using a  single frame DIT of 30 s for a total observing time of 3 hours and 20 minutes. 

\section{Direct imaging data reduction, astrometry and spectrophotometric extraction}\label{imaging_data}
The data reductions were performed with three different pipelines, namely,  the GRAPHIC pipeline \citep{Hagelberg2016}, the LAM-ADI pipeline \citep{Vigan2015,Vigan2016} and a reduction from the SPHERE Data Center (DC) using the SPHERE Data Reduction and Handling (DRH) automated pipeline \citep{Pavlov2008} for the standard cosmetic and the SpeCal pipeline for the post-processing (Galicher et al., in prep). Different types of analysis were also conducted with these pipelines, to compare and take the full advantages of each of them and extract the astrometry and spectra of HD 4747B with the best accuracy. We built the SED of the host star (Fig. \ref{sed}) based on a BT-NextGen model \citep{Allard2012}, and using VOSA \citep{Bayo2008} to extract the photometry of the primary in the different band filters and find the best fit. All three pipelines gave similar results with astrometric and photometric standard deviations respectively 66\% and 50\% smaller than the error bars given hereafter.
\begin{figure*}[!t]
\centering
	\subfloat[ $25^\circ$ double roll subtraction angle model]{\includegraphics[width=0.485\columnwidth]{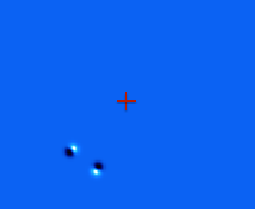}\label{NACO_model}}
\hspace{0.02 cm}
	\subfloat[$10^\circ$ double roll subtraction angle]{\includegraphics[width=0.485\columnwidth]{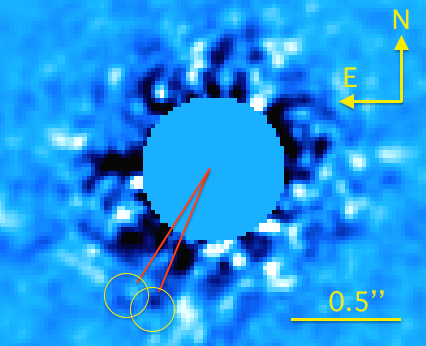}\label{NACO_10}}
\hspace{0.02 cm}
	\subfloat[$25^\circ$ double roll subtraction angle]{\includegraphics[width=0.485\columnwidth]{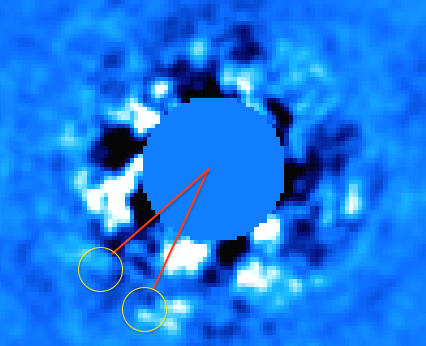}\label{NACO_25}}
\hspace{0.02 cm}
	\subfloat[Median $\chi^2$ map over double roll subtraction angles. The cross shows the minimum $\chi^2$ and so the position of the companion.]{\includegraphics[width=0.485\columnwidth]{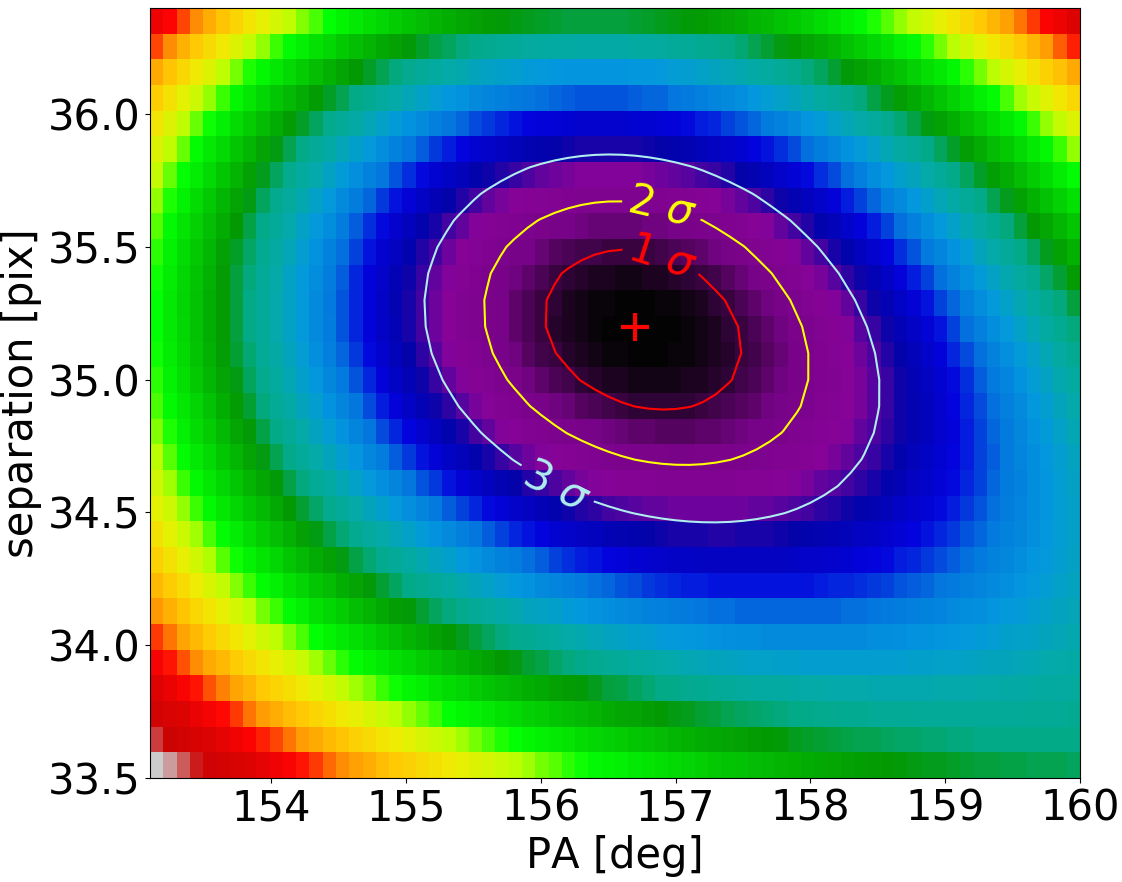}\label{med_chi2_zoo}}
\caption{NACO final images after SDI and double roll subtraction applied for different angles ($5^\circ$, $10^\circ$, and $25^\circ$), and final $\chi^2$ map of separation versus position angle.}\label{NACO_final}
\end{figure*}

\subsection{IRDIS data reduction}
\begin{figure}[!h]
\includegraphics[width=0.97\columnwidth]{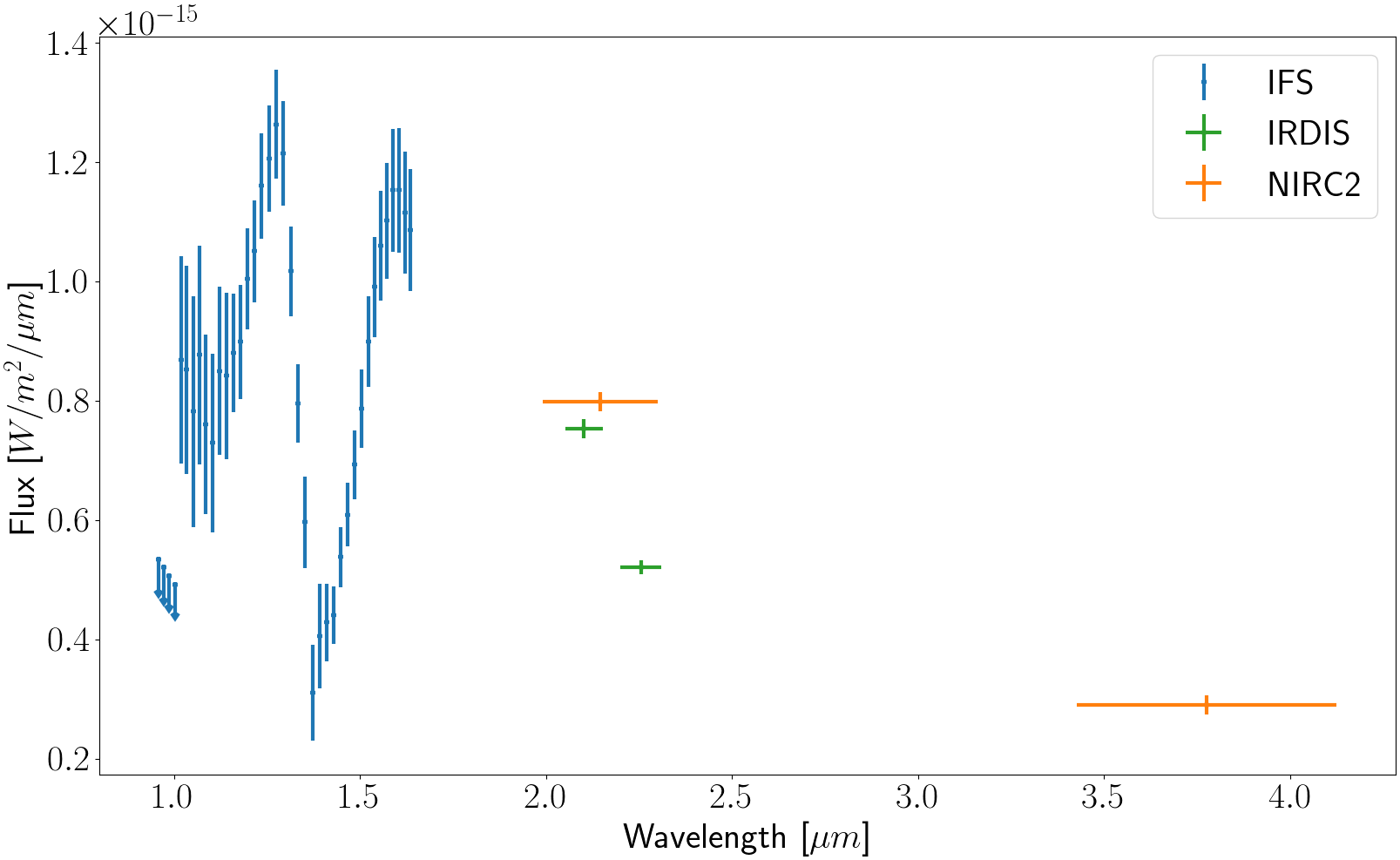}
\caption{Extracted spectrophotometry of HD 4747B. The blue points are SPHERE IFS data, the green and yellow ones are respectively IRDIS SPHERE and NIRC2 photometry. The NIRC2 data are taken from \citet{Crepp2016}. The horizontal errorbars represent the width of the IRDIS and NIRC2 filters.}
\label{spectrum}
\end{figure}

We processed the IRDIS data with several algorithms, angular differential imaging (ADI), spectral differential imaging (SDI) and radial-profile subtraction (Fig. \ref{IRDIS_reduction}). With the small separation ($\rho\simeq$600 mas) and field rotation (Sect. \ref{direct_imaging}), the ADI and SDI self-subtraction are substantial. However the companion has a small contrast with respect to its host star in K12 ($\sim$9 mag) and is visible in the raw frames. We therefore decided to use a simple median combination of the frames with a background fit (Fig. \ref{median_combine}) for the photometry extraction in $K_1$ and $K_2$ and obtained magnitude contrasts of $\Delta K_1=9.11\pm0.15$ and $\Delta K_2=9.24\pm0.15$. The error bars take into account the error on the PSF fit as well as the variations of the PSF and speckle noise through the observational sequence. We derived then from the SED of the primary its flux in each filter, and finally extracted the flux of HD 4747B (Fig. \ref{spectrum}).

The astrometric extraction was done using a simple radial-profile subtraction (Fig. \ref{radial_profile}), as it allows to increase the signal to noise ratio and it does not strongly affect the shape of the companion point spread function (PSF). The derived separations and position angles are listed in Table \ref{astrometry}. The error bars given in the table include astrometric calibration errors that were quadratically added to the results of the adjustment.

\subsection{IFS data reduction and spectral extraction}
We used a TLOCI-ADI reduction to extract the IFS photometry \citep[][Specal implementation used]{Marois2014}, which was optimized for getting the best contrast in each wavelength with the least possible impact on the spectrum of the companion. The contrast in each wavelength channel is computed by injecting fake companions before the ADI processing. This allows to account for the self-subtraction effect from the ADI reduction. The apparent spectrum of the companion is then extracted with the same procedure as for the IRDIS data, by computing the primary flux in each channel given its SED. The extracted spectrum is shown in Fig. \ref{spectrum}.
\subsection{NACO data reduction and astrometric extraction} \label{NACO_reduc}
We reduced the NACO data following the standard data processing, bad-pixel cleaning, subtraction of sky and division by flats. The SDI sub-images were extracted and re-centered by adjusting a simple Gaussian profile to the wings of the primary star. We performed a frame selection based on the integrated flux of each image. Images with a flux higher than the median value are rejected, which allows to remove frame taken with lower seeing conditions and/or AO performances. The shorter wavelength images are resized by a factor $\lambda1/\lambda2$ and centered by the cross-correlation to the longer wavelength images. For each time step, the images taken at different wavelengths are subtracted. Images with the same derotator angle are co-added using their median value. 

Finally, different sets of double roll subtractions are performed to remove residual pupils aberrations and to amplify the companion signature. For instance, we subtracted every image pair that have a pupil rotation offset of $10^\circ$, {\it ie.}  ${\rm Im}(10^\circ)-{\rm Im}(0^\circ)$, ${\rm Im}(15^\circ)-{\rm Im}(5^\circ)$, ... , ${\rm Im}(95^\circ)-{\rm Im}(85^\circ)$ and median combined them. We did the same on image pairs with  $5^\circ$,  $15^\circ$, $20^\circ$, and  $25^\circ$ pupil rotation offsets.
This data analysis technique affects the companion signature in the final image, which is composed of 4 duplicated PSF as shown on Fig. \ref{NACO_final}a. The SDI part of the algorithm results in the radial positive/negative part of the companion signature, while the double roll subtraction duplicates the SDI pattern with the chosen rotation offset. We found that the $10^\circ$ pupil rotation offset performs the best in term of noise reduction and signal amplification (see Fig.~\ref{NACO_final}b) which allow us to detect HD 4747B in the archived NACO data.
We retrieved the companion astrometry and corresponding confidence intervals by modelling the expected companion pattern and computing a $\chi^2$ map for different companion's separations, position angles and flux ratio. A clear minimum in the $\chi^2$ map is seen at the expected position of the companion as illustrated on Fig. \ref{med_chi2_zoo}. Each individual full $\chi^2$ maps for the different double roll subtraction angles are shown in Fig.~\ref{chi2_all}, while Fig. \ref{chi2} and Fig. \ref{med_chi2_zoo} show the median (See Appendix \ref{naco_appendix} for more details about this analysis).

\begin{figure}[!t]
\includegraphics[width=0.97\columnwidth]{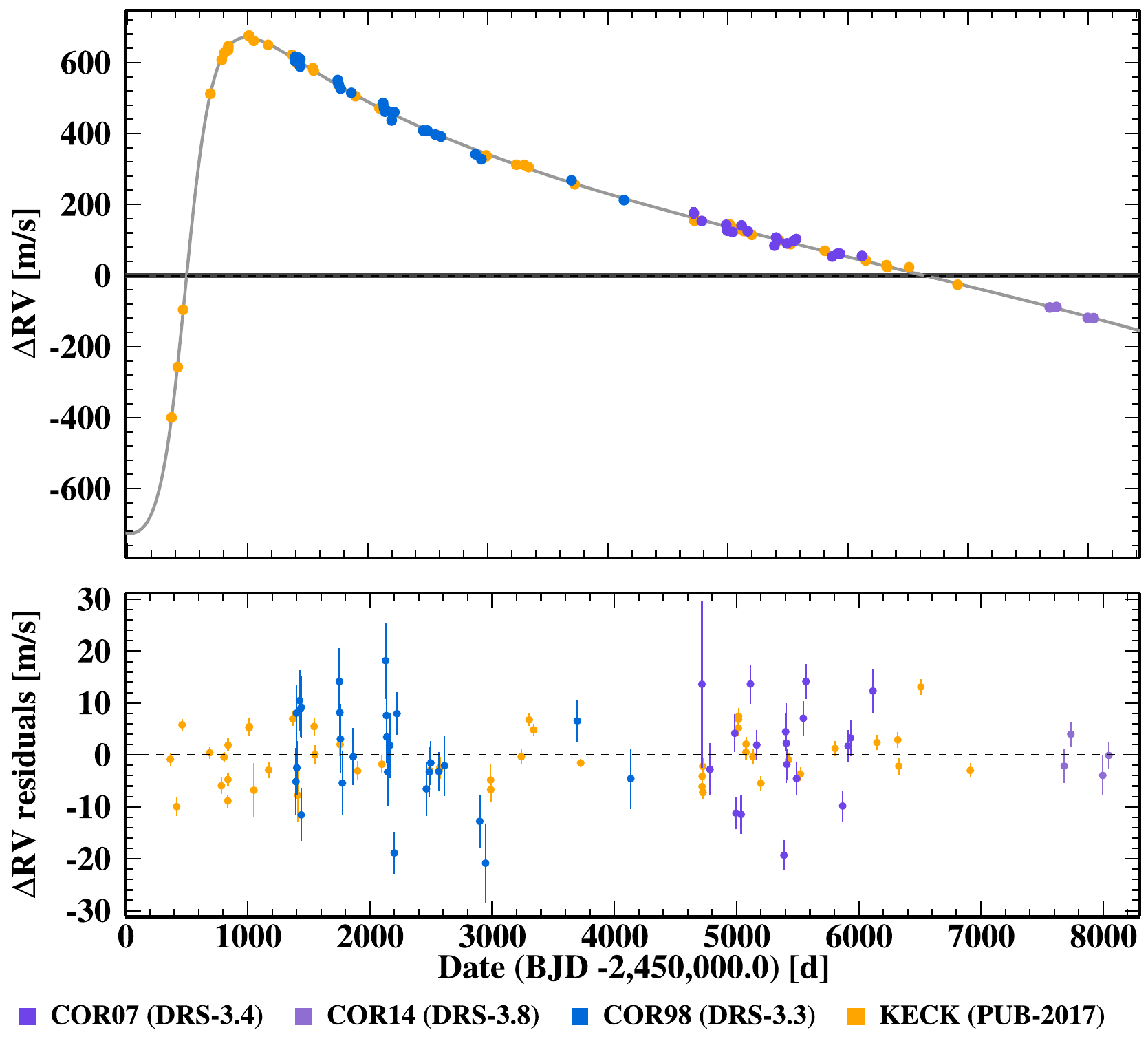}
\caption{Radial Velocity measurements for HD 4747 taken with the Keck-HIRES and CORALIE spectrographs. The different versions of CORALIE are indicated with different colors. The best fit model from the combined MCMC analysis with direct imaging is marked with the grey line.}
\label{RV_curve}
\end{figure}

\begin{figure}[!t]
\includegraphics[width=0.97\columnwidth]{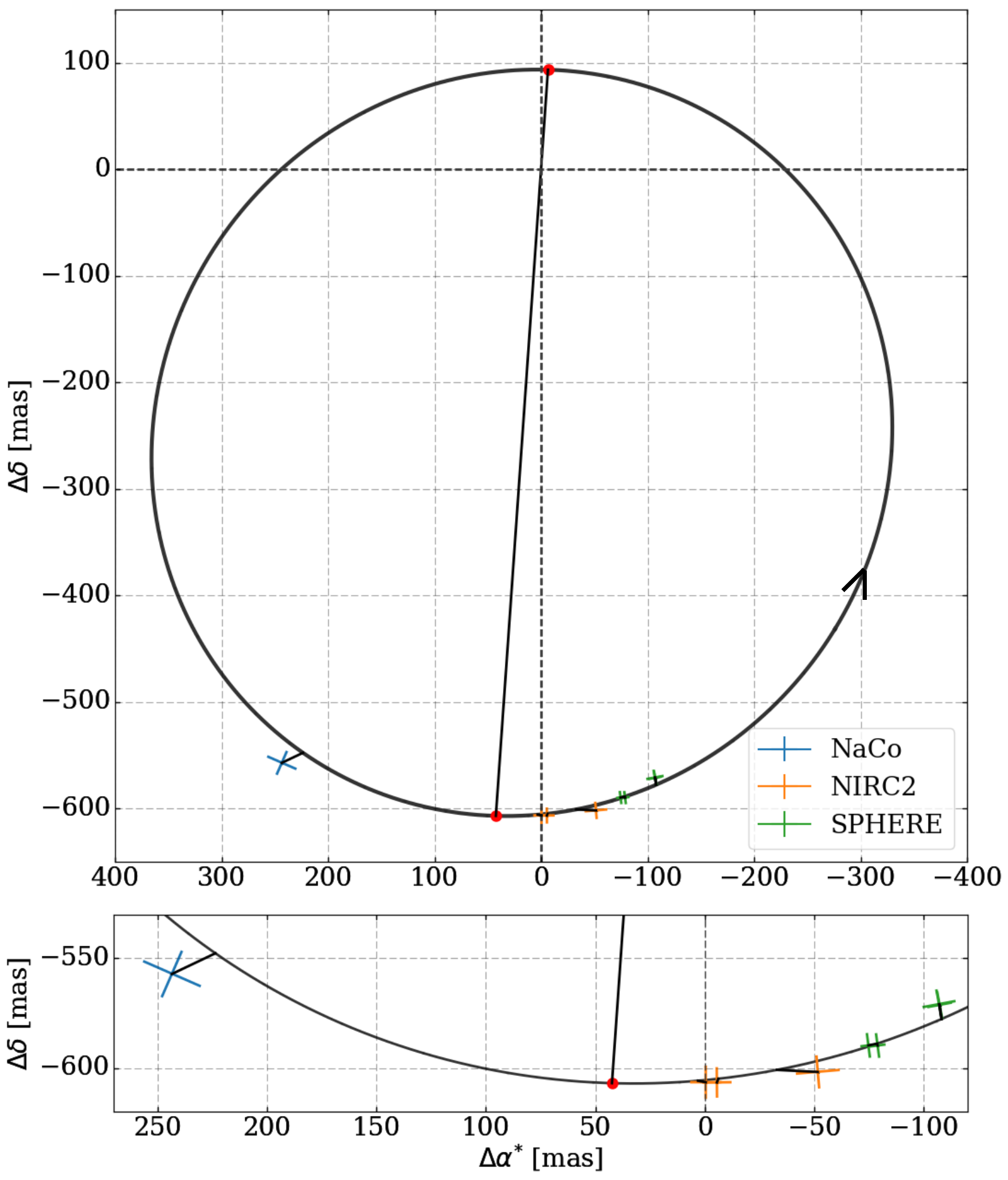}
\caption{Relative orbit of the HD 4747AB system. The black curve corresponds to the maximum likelihood of the combined RV and direct imaging MCMC analysis. The NaCo data point is shown in blue, the NIRC2 $K_{s}$ and $L'$ observations from \citet{Crepp2016} in yellow, and the SPHERE ones in green. The arrow shows the orbit rotational direction and the red dots show the apastron and periastron.}
\label{visual_orbit}
\end{figure}

Deriving reliable confidence intervals for these archived data is also a challenging task. We computed the noise of the image for each separation using the standard deviation in annulus of 1 pixel width. This noise map was used to compute the $\chi^2$ map and the 1,2 and 3 $\sigma$ confidence intervals (see Fig.~\ref{NACO_final}d). As the center of the star is difficult to determine behind the coronagraph, we decided to add quadratically a systematic error of 4 mas on the separation measurement, which corresponds to a quarter of a pixel.
As no astrometric field was observed during this run with NACO, and no calibration could be found for the SDI+4 mode's pixel scale, we took the NACO manual value of 17.32 mas/px. We quadratically added an error of 0.05 mas/pixel to the budget value, which corresponds to an error of 1.8 mas at the separation of the companion. For the True North, we used the calibration from \citet{Ehrenreich2010} that was taken on the $20^{th}$ of august 2008, 18 days before our dataset. The true north of NACO was at a $0^\circ\pm0.2^\circ$ angle. We quadratically added an uncertainty of $0.5^\circ$ as the detector could have moved slightly between the runs.

At the end, an angular separation of $\rho=608\pm11$ mas and a position angle of $PA=156.4^\circ\pm1.3^\circ$ are retrieved.

We decided to not use the photometry of this data set. The detection is indeed at the noise limit, and the flux strongly polluted by the self-subtraction of both SDI and the double roll subtraction. Moreover, the filters are redundant with the SPHERE IFS channels and no PSF calibration flux were correctly done.

\begin{table}[!h]
\centering
\captionof{table}{Measured astrometry and contrast of HD 4747B from SPHERE ($K_1$ and $K_2$) and NACO (H, SDI mode) data.}\label{astrometry}
\begin{small}
\begin{tabular}{ccccc}
\hline
\hline
Filter  & Date (UT) & $\rho$ (mas)  & $PA$ (deg) & Contr. (mag) \\
\hline
$K_1$ & 12.12.2016 & 594.4 $\pm$ 5.1 & 187.2 $\pm$ 0.3 & $9.11 \pm 0.15$\\
$K_2$ & 12.12.2016 & 595.0 $\pm$ 5.1 & $187.6 \pm 0.3$ & $9.24 \pm 0.15$\\
$K_1$ & 28.09.2017 & 581.2 $\pm$ 5.8 & 190.6 $\pm$ 0.5 & -\\
$K_2$ & 28.09.2017 & 580.8 $\pm$ 6.3 & $190.6 \pm 0.7$ &  -\\
$H$ & 07.09.2008 & 608 $\pm$ 11 & 156.4 $\pm$ 1.3 & -\\
\hline
\hline
\end{tabular} 
\end{small}
\end{table}

\section{Orbital analysis and dynamical mass estimation}\label{orbit}
We performed an MCMC analysis combining all radial velocity data available (see Sect. \ref{RV_sect} and Fig. \ref{RV_curve}) with our direct imaging SPHERE epochs (see Sect. \ref{imaging_data} and Fig. \ref{visual_orbit}), the epochs from \citet{Crepp2016} and the NACO SDI+4 archival data point (See Sect. \ref{NACO_reduc}). This epoch allows us to constrain strongly the orbital parameters thanks to the much longer time baseline. The MCMC simulation was performed by using $emcee$ \citep{Foreman2013}, a python implementation of the affine-invariant ensemble sampler for MCMC proposed by \citet{Goodman2010}. The data are modeled with a Keplerian and 4 RV offsets (one for HIRES, and 3 for the different versions of CORALIE: C98, C07, C14). The noise in the radial velocity data is modeled with a nuisance parameter for each instrument. As we have direct imaging and radial velocity measurements, the parallax and mass of the primary are parameters that can be constrained with the fitting. We introduce these two parameters with Gaussian priors taken from Table \ref{t:param}. More details about the MCMC orbital analysis are presented in Appendix \ref{orbital_mcmc_appendix}. The results of the orbital fit are presented in Table \ref{orbital_parameters}, Fig. \ref{triangle_DV} and \ref{triangle_orbit}.

\begin{table}[!htb]
\centering
\captionof{table}{Orbital parameters from the maximum likelihood}\label{orbital_parameters}
\begin{tabular}{cc}
\hline
\hline
Parameters & Values ($1\sigma$) \\
\hline
P [yr] & $33.08\pm0.70$\\
K [m/s] & $698.0\pm10.4$ \\
ecc & $0.7320 \pm0.0023$ \\
$\omega$ [deg] & $-93.10 \pm 0.47$ \\
$T_0$ [bjd] & $50473.9\pm5.2$\\
$\Omega$ [deg] & $89.9\pm1.4$\\
$i$ [deg] & $46.3\pm1.1$\\
$m sin(i)$ [$\rm M_{Jup}$] & $50.7\pm1.8$\\
\hline
m$_1$ + m$_2$ [$\rm M_\odot$] & $0.918\pm0.037$\\
m$_2$ [$\rm M_{Jup}$] & $70.2\pm1.6$\\
a [au] & $10.01\pm0.21$\\
\hline
\hline
\end{tabular} 
\end{table}

Compared to the orbital solutions found by \citet{Crepp2016}, we find significant differences in the results of our MCMC analysis. The period $P=33.08\pm0.70$ yr is in good agreement with \citet{Sahlmann2011}, but disagrees with the \citet{Crepp2016} value at $6\sigma$. The high eccentricity predicted in \citet{Sahlmann2011} and confirmed by \citet{Crepp2016} is verified, $e=0.7320\pm0.0023$, even if our value is slightly smaller. The inclination of $46.3^\circ\pm1.1^\circ$ derived leads to a semi-major axis of $a=10.01\pm0.21$ au and a mass estimation for HD 4747B of $m_{\rm B}=70.2\pm1.6M_\odot$. This inclination is smaller by $3\sigma$ than the value from \citet{Crepp2016}. This explains the difference in the mass measurement.  The discrepancy between our orbital parameters and those of \citet{Crepp2016} likely arises from the Keck 2015 L' band measurement. This datapoint is off by more than $1\sigma$ and with more epochs, our fit is less sensitive to individual outliers.

\section{Spectrophotometric analysis} \label{spec_phot}
\subsection{Color-magnitude diagram}
\begin{figure}[!h]
\includegraphics[width=0.97\columnwidth]{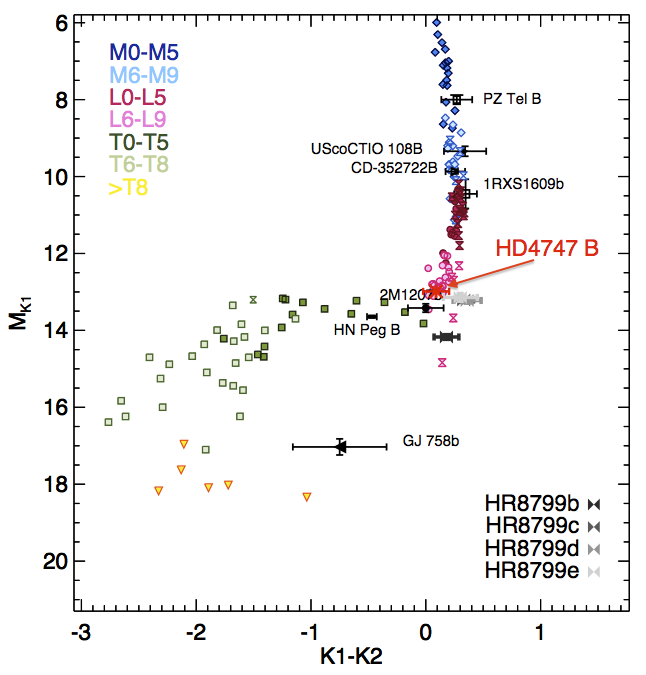}
\caption{Color-magnitude diagram of HD 4747B from the IRDIS K1 and K2 observations. Other known objects are shown for comparison.}
\label{color_mag}
\end{figure}

From the IRDIS K1 and K2 observations the color-magnitude diagram of HD 4747B shows a late L spectral type (Fig. \ref{color_mag}). This is in good agreement with the prediction of \citet{Crepp2016}, which used the broad bands Ks and L' filters from the NIRC2 camera at the Keck Telescope. HD 4747B falls close to the HR8799 c,d and e planets on the diagram and at the L-T transition.

\subsection{Known standards comparison}
\begin{figure}[!b]
\includegraphics[width=0.97\columnwidth]{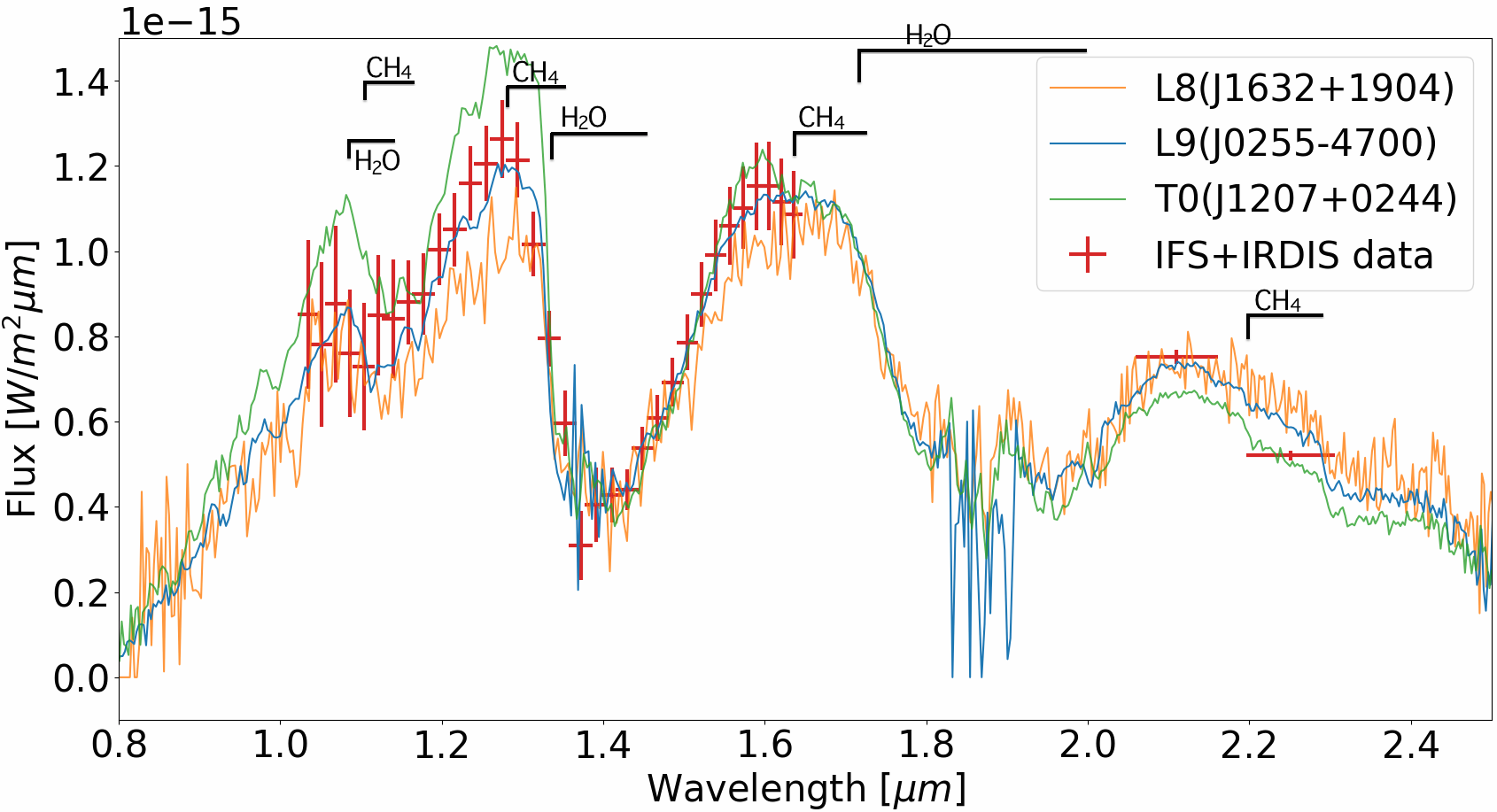}
\caption{Adjustments of standard known objects showing L8 \citep[J16322911+1904407;][]{Burgasser2007}, L9 \citep[J02550357-4700509;][]{Burgasser2006} and T1 \citep[J12074717+0244249;][]{Looper2007}. The best fit is done by the L9 object. The molecular absorption bands of CH$_4$ and H$_2$O are indicated.}
\label{fit_spectrum}
\end{figure}
To further constrain the spectral type of HD 4747B, a comparison of standard near-IR spectra of L1 to T8 known objects from the SpeX Prism Library \citep{Burgasser2014} was done. To do so we made use of the $splat$ python package \citep{Burgasser2016}. Each standard spectrum is first normalized in flux and then calibrated to match the extracted data of HD 4747B. The fitting of each standard spectrum ($F_k$) on the one of HD 4747B ($f$) is done by comparing the goodness of fit:

\begin{equation}
\chi^2=\sum^n_{i=1} \left(\frac{f_i-C_kF_{k,i}}{\sigma_i}\right)^2
\end{equation}
where $C_k$ is the flux scaling factor. Each standard spectrum was binned to the appropriate spectral resolution of the IFS measurements with a Gaussian convolution. We decided to use a FWHM of 1.5 times the separation between each wavelength in order to take into account the correlation between the IFS channels. The IRDIS fluxes were estimated by using the transmission curves of the K1 and K2 filters. The L' NIRC2 observation was not used as the spectra from this library stop in the K band. Fig. \ref{fit_spectrum} shows the fitting results for the L8, L9 and T0 standards. The best fit corresponds to the L9 dwarf DENIS-P J0255-4700 \citep{Burgasser2006} and matches the IFS and IRDIS data well. This is in good agreement with the color-magnitude diagram (Fig. \ref{color_mag}). \citet{Crepp2016} derived also a late L type but a higher effective temperature of $T_{\rm eff}=1700\pm100$ K. This led them to a model dependent mass higher than in this study.

\subsection{Atmospheric forward modeling: Exo-REM}\label{Exo-REM_section}
We characterized the observations with the forward model Exo-REM \citep{Baudino2015,Baudino2017} using grids of synthetic models generated with a $T_{\rm eff}$ between 400 and 1800 K (with a step of 50 K), a log($g$[cgs]) between 2.5 and 5.5 (with a step of 0.5), a diversity of clouds \citep[without cloud, or with $\tau_{\rm ref}$=0.5 or 1, see in ][]{Baudino2015}, a metallicity z = -0.2, 0, 0.5, 1.0 and taking account of the equilibrium chemistry or non-equilibrium chemistry (with a $k_{\rm zz} = 10^{8}$ $\rm cm^2 s^{-1}$, see \citet{Baudino2017}, for the non-equilibrium chemistry formalism). The parameters of the best model is summarized in Table \ref{exo_rem} and the best fit spectra are shown Fig. \ref{Exo_Rem_best_fit}. The maximum mass at $1\sigma$ computed from the gravity and radius derived with Exo-REM is only 14.6 $M_{\rm Jup}$, which is far below the dynamical mass measurement. The difference between the results with Exo-REM and the dynamical mass can be explained by the difficulty to generate the more extreme surface gravity with this model (built for giant planets, i.e. low gravity objects). HD 4747B is indeed the most massive brown dwarf studied with Exo-REM and is at the gravity limit available with this model.
\begin{table}[!h]
\centering
\captionof{table}{Results of the forward modeling Exo-REM at $5\sigma$ detection for HD 4747B}\label{exo_rem}
\begin{small}
\begin{tabular}{ll}
\hline
\hline
Parameter & Values\\
\hline
$T_{\rm eff}$ (K) & $1300\pm100$\\
log($g$) (dex) & $4\pm0.5$\\
Cloud condition & cloudy\\
Chemistry & no conclusion\\
Radius ($\rm R_{Jup}$) & $0.91\pm0.16$\\
Metallicity (solar) & 0.63-1\\
\hline
\hline
\end{tabular} 
\end{small}
\end{table}

\begin{table*}[!t]
\centering
\caption{Parameters and Priors Used in the Retrieval analysis}\label{tab:priors}
\begin{tabular}{llll}
\hline
\hline
Parameter & Symbol & Prior Used & Value \\
\hline
Radius & $R$ & Gaussian & $R_{\rm comp} = 1.0 \pm 0.1 $  $\rm R_{Jup}$ \\
Planet mass & $M_{\rm comp}$ & Gaussian & $70.2 \pm 1.6$ $\rm M_{Jup}$ \\
Molecules abundances or elemental abundances & $X_i, f_i$ & Log-uniform & $10^{-15}$ to $10^{-1}$ \\
Longwave/infrared opacity (TP profile) & $\kappa_0$ & Log-uniform & $\log{\kappa_0} = 10^{-15}$ to 10 (mks) \\
Internal/interior temperature (TP profile)  & $T_{\rm int}$ & Uniform & 100 to 2300 K \\
Extinction coefficient & $Q_0$ & Uniform & 1 to 100 \\
Cloud particles size & $r_{\rm c}$ & Log-uniform & $10^{-7}$ to $10^{-3}$ m \\
Cloud particle abundance & $f_{\rm cloud}$ & Log-uniform & $10^{-30}$ to $10^{-4}$ \\
Distance & $d$ & Gaussian & $19.25 \pm 0.58$ pc \\
\hline
\hline
\end{tabular}
\end{table*}
\subsection{Atmospheric retrieval modeling: HELIOS-R}
For this analysis we used the atmospheric retrieval code HELIOS-R developed by \citet{lavie17}. Atmospheric retrieval is a technique borrowed from the Earth remote sensing community.  Some pieces of the atmospheric physical model are parametrised (i.e. Temperature-pressure profile, clouds etc.). It sacrifices self-consistency in order to speed up computational time, which in return allows for a more robust parameter space exploration and a better characterization of the uncertainties on the model parameters.

HELIOS-R allows for a direct comparison of different 1-dimensional emission forward model using the Nested Sampling algorithm \citep{skilling06}. The model parameters and their priors are presented in Table \ref{tab:priors}. As in \cite{lavie17}, we assume independent Gaussian errors so the likelihood takes the form of equation \ref{likelihood_HELIOS}:
\begin{equation}
\mathcal{L}(D|M_i,\theta) = \prod_{k=1}^N \frac{1}{\sigma_k\sqrt{2\pi}}exp\left(-\frac{[D_{k,obs}-D_{k,model}]^2}{2\sigma_k^2}\right),\label{likelihood_HELIOS}
\end{equation}
where $D_{k,obs}$ is the k-th observational data point, $D_{k,model}(\theta)$ the model prediction for this data point given the parameters $\theta$, $\sigma_k$ the uncertainty of the k-th observational data point, and N the total number of data points. This approach does not account for covariances in the IFS data.

The models assume an Hydrogen/ Helium dominated atmosphere and include the four main absorbers in the infrared: carbon monoxide (CO), carbon dioxide (CO$_2$), water (H$_2$O) and methane (CH$_4$). Two sets of assumptions can be made regarding the atmosphere chemistry : equilibrium chemistry where the two parameters are the carbon ($f_c$) and oxygen ($f_o$) abundances; and unconstrained chemistry where the parameters are the four molecules abundances assumed to be constant throughout the vertical 1D atmosphere. Clouds are modeled using a three parameter model, first introduced in \cite{lee13}. See \cite{lavie17} in order to get more insight on HELIOS-R.

\subsubsection{Companion mass, gravity and radius priors}
As discussed in \citet{lavie17}, retrieving the surface gravity and radius of directly imaged objects is challenging. With the current number of atmospheric data (spectrum and photometry) and their uncertainties, it is impossible with an atmospheric retrieval technique to constrain the surface gravity at the same level of precision as with the radial velocity data. In the case of HD 4747B, velocity data are available. This valuable information need to be taken into account in the atmospheric Bayesian analysis. We have updated HELIOS-R in order to take the companion mass as a parameter of the model. The mass, surface gravity and radius are linked by the following equation :
\begin{equation}
M_{\rm comp} = g * R^2_{\rm comp} / G \label{eq:mass}
\end{equation}

The companion mass constrained by the radial velocity can now be enforced in our prior in a straightforward manner. 
There is no direct measurement of the radius on this object as it does not transit. Consequently, our prior should reflect our current state of knowledge on the radius of brown dwarfs. The evolutionary track \citep[e.g.][]{Baraffe2003} gives a radius of ~ 0.9 R$_{\rm Jup}$ for HD 4747B. However, there is not a unique mass-radius relationship for a given brown dwarf depending on what one assumes in the models \citep{Burrows:2011aa}. \cite{Konopacky:2010aa} showed as well that traditional evolutionary tracks are missing physics or chemistry. Observations of transiting brown dwarfs show that those objects have radii from $\sim$0.8 to 1.2 R$_{\rm Jup}$. We therefore set our prior as a Gaussian prior of R = 1.0 $\pm$ 0.1 R$_{\rm Jup}$\footnote{This gaussian prior takes into account the 0.8 to 1.2 R$_{\rm Jup}$ distribution at 2$\sigma$.}.

\subsubsection{Results and discussion}\label{HELIOS_R_results}
\begin{figure}[!h]
\centering
\includegraphics[width=0.97\columnwidth]{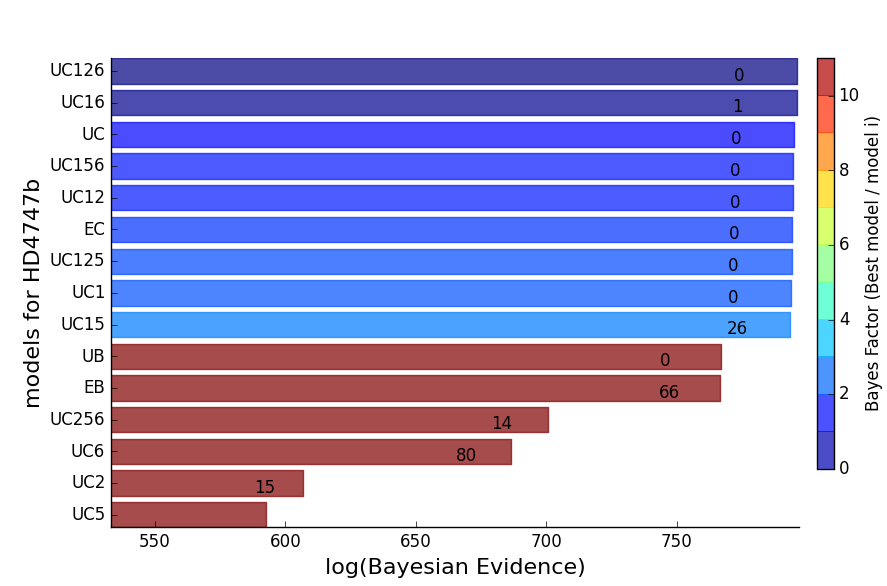}
\caption{Bayes factors from a suite of models for HD 4747B. Four models were considered: equilibrium chemistry without cloud (EB) or with clouds (EC) and unconstrained chemistry without (UB) or with clouds (UC). The numbers associated with each model's name indicate the molecules included (1: H$_2$O, 2: CO$_2$, 5: CO and 6: CH$_4$); if there are no numbers then all four molecules are included. The Bayesian evidence clearly favors models with clouds and the most favored model is unconstrained chemistry including H$_2$O, CO$_2$ and CH$_4$. The number associated with each histogram is the logarithm of the Bayes factor between the model in question and its neighbor below.  The color bar shows the logarithm of the Bayes factor between the model in question and the best model, which is the model placed at the top.}\label{fig:evidence}
\end{figure}
 \begin{table}[!h]
\centering
\caption{Summary of Retrieved Results.}\label{tab:results}
\begin{tabular}{lc}
\hline
\hline
Property  & Value \\
\hline
$X_{\rm H_2O}$  & $-3.57_{-0.07}^{+0.07}$   \\ 
$X_{\rm CO_2}$  & $-6.13_{-5.11}^{+1.92}$   \\ 
$X_{\rm CO}$  & $-9.65_{-3.08}^{+3.28}$   \\ 
$X_{\rm CH_4}$  & $-4.62_{-0.43}^{+0.25}$   \\ 
$Q_0$  & $0.88_{-0.50}^{+0.58}$   \\ 
$r_{c}$ [m]  & $-4.57_{-0.73}^{+0.86}$   \\ 
$X_{c}$  & $-21.00_{-1.56}^{+1.32}$   \\ 
$d$ [pc] & $19.56_{-0.41}^{+0.35}$   \\ 
$M_{\rm p}$ [M$_{\rm Jup}$]  & $70.09_{-1.21}^{+1.25}$   \\ 
$R_{\rm p}$ [R$_{\rm Jup}$]  & $0.85_{-0.03}^{+0.03}$   \\ 
\hline
$\mu$  & $2.20_{-0.00}^{+0.00}$   \\ 
$C/O$  & $0.13_{-0.08}^{+0.14}$   \\ 
$C/H$  & $-4.72_{-0.39}^{+0.47}$   \\ 
$O/H$  & $-3.79_{-0.08}^{+0.16}$   \\ 
$\log{g}$ [cgs]  & $5.40_{-0.03}^{+0.03}$   \\ 
\hline
\hline
\end{tabular}
\tablefoot{We have listed the $1\sigma$ uncertainties, which were computed by locating the 15.87th and 84.13th percentile points on the horizontal axis. In the limit of a symmetric Gaussian function, these would yield to its fwhm. The mean molecular weight is given at 1 bar. Values are in $\log_{10}$ (except for C/O, d, $M_{\rm p}$,$R_{\rm p}$ and $\mu$) and dimensionless (except when units are shown).}
\end{table}

\begin{figure}[!h]
\centering
\includegraphics[width=0.97\columnwidth]{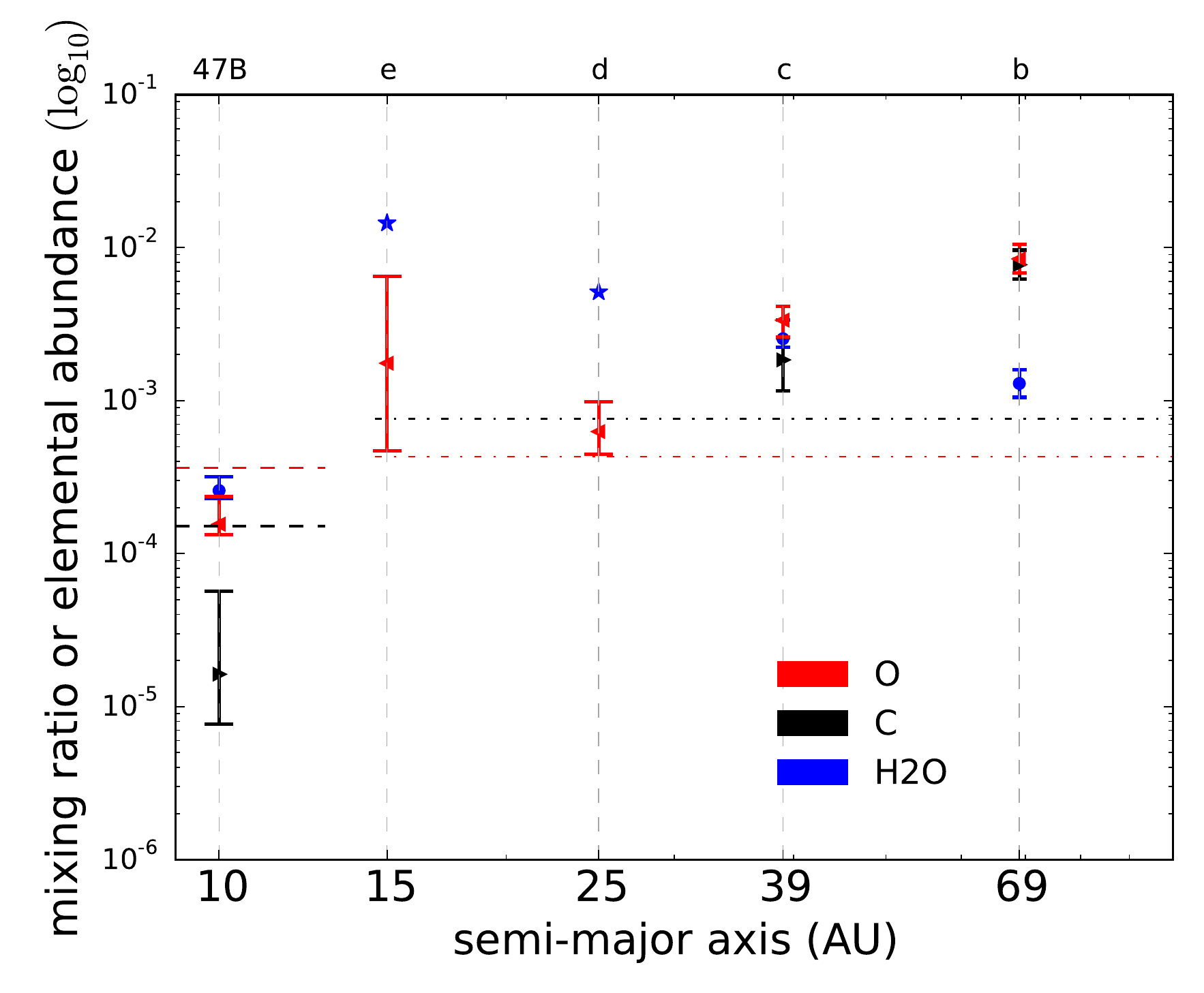}
\includegraphics[width=0.97\columnwidth]{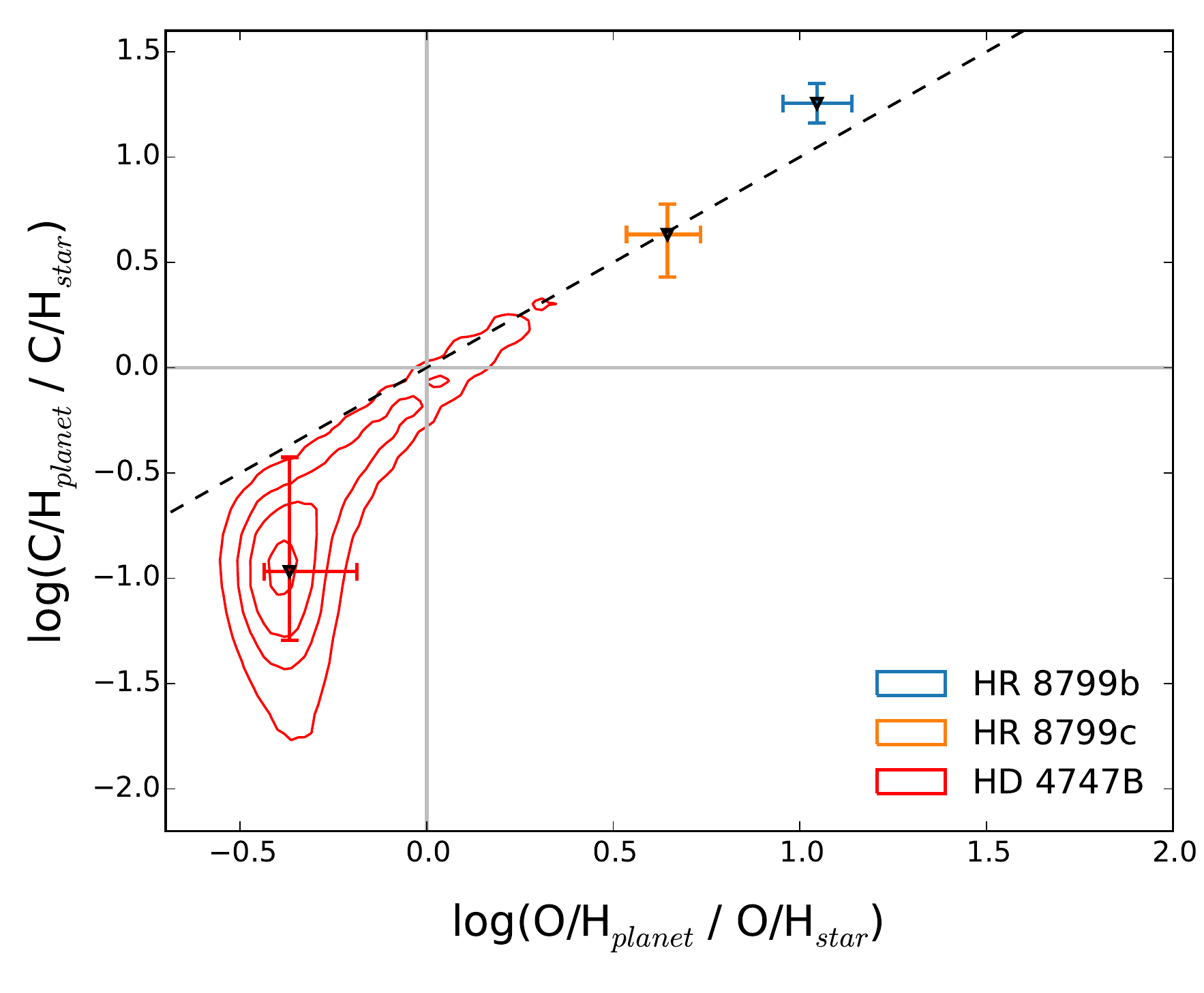}
\caption{Results from the atmospheric retrieval analysis. The top panel shows the retrieved water mixing ratios and elemental abundances of carbon and oxygen for HD 4747B as well as the four HR 8799 exoplanets as a function of the distance to the host star. For HR 8799d and e, we show the water abundance in chemical equilibrium at 1 bar (represented by the blue stars). The carbon abundance retrieved for these two planets is very small and is not shown on this figure, see Fig \ref{fig:summary_2}. The carbon and oxygen abundances of the stars are shown with the dashed lines (HD 4747) and dashed dot lines (HR 8799).
The bottom panel shows the companion elemental abundances normalized to its stellar values with the dashed black line denoting parity.}\label{fig:summary}
\end{figure}

The main output of the Nested Sampling algorithm is the Bayesian evidence, which allows the models comparison through the computation of the Bayes factor (${\cal B}$). We considered four different models : equilibrium chemistry without cloud (EB) or with clouds (EC) and unconstrained chemistry without (UB) or with clouds (UC). For the models with unconstrained chemistry, we also considered all the different combinations of molecules in order to evaluate the significant of each molecules detection. Fig. \ref{fig:evidence} shows the Bayesian evidence for most of the models considered. Models with clouds are strongly preferred over the models without clouds, which indicates that this object is probably cloudy. The unconstrained chemistry model with clouds is the preferred model. The Bayes factor between this model (including all the molecules) and the models excluding some molecules indicate that CO$_2$ and CH$_4$ are only weakly detected ($\ln{\cal B} < 2.5$), while H$_2$O is strongly detected ($\ln{\cal B} > 5$). We have a low sensitivity to CO, since including it or not in the model does not change the constraints on the other parameters and does not improve the fit to the data. See also \cite{Trotta:2008aa} for a correspondence between the Bayes factor ({\cal B}) and the significance in terms of the number of standard deviations.
A summary of the retrieved parameters for the UC model (including all the molecules) is shown in Table \ref{tab:results}, the best fit spectrum and temperature profile are shown in Fig. \ref{fig:spectra} and the posterior distribution of each parameters are shown in Fig. \ref{fig:retrieval1}.

The dimensionless cloud parameter $Q_0$ serves as a proxy for the cloud composition. It is not possible to determine the composition as the posterior distribution is too broad and encompasses the refractory species composition (e.g., silicates - $Q_0 \sim 10$) and the volatile species composition (e.g., ammonia, methane, water - $Q_0 \sim$ 40-80). However, the posterior distribution of the clouds particle size indicate that the cloud is composed of big particles, which will act as a constant absorbers as a function of the wavelength \citep{lavie17}.

The retrieved posterior distribution of the companion mass reflects our prior, which is normal as the radial velocity data (used to build our prior) provide a better constrain on this parameter than the spectrum.  With the retrieved companion radius ( $0.85$ $\rm R_{Jup}$ ) and mass we are able to compute the surface gravity of the companion using Eq. \ref{eq:mass}, which indicates a high gravity object ($\log{g} = 5.40$ cgs).

We have measured oxygen abundances by exploiting the O I triplet at 7771-7775 Angstroms, and applying 3D NLTE corrections by \citet{Amarsi2015}, as done in \citet{dorazi2017}. We have found a slight over-abundance being [O/Fe]=+0.16$\pm$0.08 dex. Considering our carbon abundance estimation given in Sect. \ref{star}, we obtained C/O=0.417. Using the retrieved molecular abundances, it is possible to derive the carbon to oxygen (C/O), the carbon to hydrogen (C/H) and the oxygen to hydrogen (O/H) ratios. The retrieved values are shown in Table \ref{tab:results}. Fig. \ref{fig:summary} shows a comparison of those values with the HR 8799 system from \cite{lavie17}, where the data comes from \citet{Bonnefoy2016,Zurlo2016}. Those values can be compared to the stellar abundance, which are 10$^{-3.82}$ and 10$^{-3.44}$ for carbon and oxygen respectively, derived from the results of Section \ref{star}. 

The interest for C/O, C/H and O/H ratios lie in the implications for planet formation. \citet{oberg11} previously outlined a first order scenario based on the position of the different snow lines. The core accretion scenario is a multi-step process that will result in a broad-range elemental composition depending on the position of the object during those different steps. The HR 8799 planets are compatible with such a scenario.  On the other hand a gravitational instability scenario is a quick one step process, which will form a companion with a similar composition to the host star formed from the same protoplanetary disk. Our retrieved values for HD 4747B are compatible within one sigma to the host star, which indicates that both scenarios are possible. However, the relative low mass of the star and the companion high mass indicate that a core accretion scenario would be difficult. It is therefore very likely that this system has formed like a binary system. More observations, especially spectroscopical one in the K-band to constrain the carbon abundance \citep{lavie17}, are requested to confirm this scenario through the atmospheric retrieval analysis. 

\subsection{BT-Settl model comparison}
\begin{figure}[!htb]
\includegraphics[width=0.97\columnwidth]{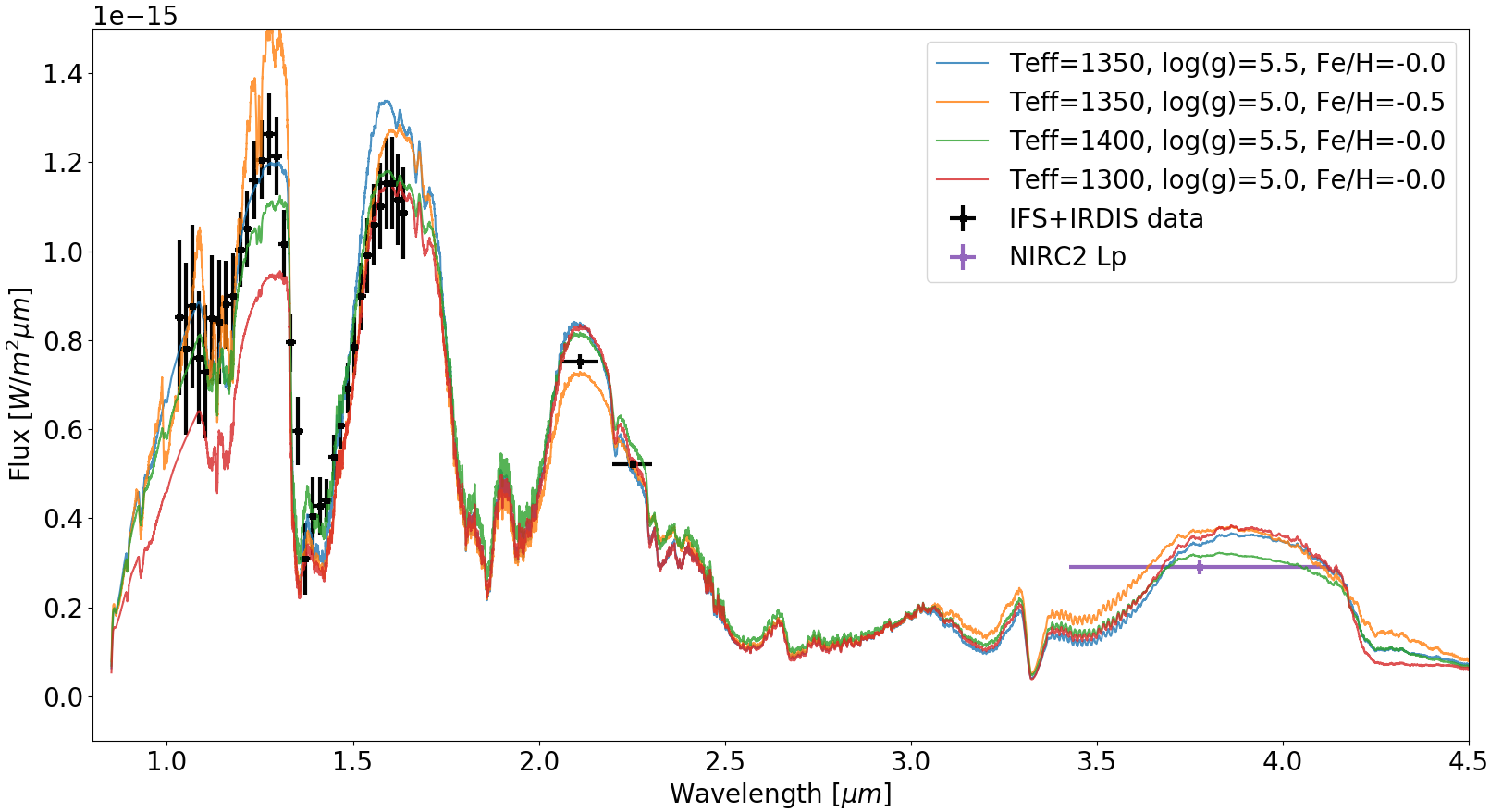}
\caption{Four best-fit spectra from the BT-Settl models \citep{Allard2011} in order from the top (best) to the bottom (less good). The adjustment was done on a grid of $T_{\rm eff}$ from 1200 K to 1700 K, with a log($g$) from 4.0 to 5.5 and a metallicity of +0.5, 0.0 or -0.5. The results are coherent with a $T_{\rm eff}=1300-1400$ K, $\log(g)=5-5.5$ and [Fe/H]$=0.0/-0.5$.}\label{Baraffe_fit_spectra}
\end{figure}

To compare the Exo-REM forward modeling and HELIOS-R retrieval with evolutionary models, we used the BT-Settl atmospheric models \citep{Allard2011}, that are combined with the BHAC15 evolutionary tracks \citep{Baraffe2015}. This model is well suited to analyse objects that are at the L-T transition such as HD 4747B. We compared our extracted spectrum to model spectra with $T_{\rm eff}$ from 1200 K to 1700 K (with steps of 50 K), with a log($g$) from 4.0 to 5.5 (steps of 0.5) and a metallicity of +0.5, 0.0 or -0.5. The four best fitting spectra (binned by a factor of 1000) are shown in Fig. \ref{Baraffe_fit_spectra}. The results point towards a temperature $T_{\rm eff}=1300-1400$ K, $\log(g)=5.0-5.5$ and a metallicity between 0.0 and -0.5. The temperature agrees well with the result from the standard-objects fitting and Exo-REM forward modeling. The gravity is in good agreement with the HELIOS-R retrieval code. The metallicity is in good agreement with the metallicity of the host star ([Fe/H]$=-0.23\pm0.05$, see Table \ref{t:param}), even if a thinner grid would be needed to constrain correctly this parameter. 

\begin{figure}[!htb]
\includegraphics[width=0.97\columnwidth]{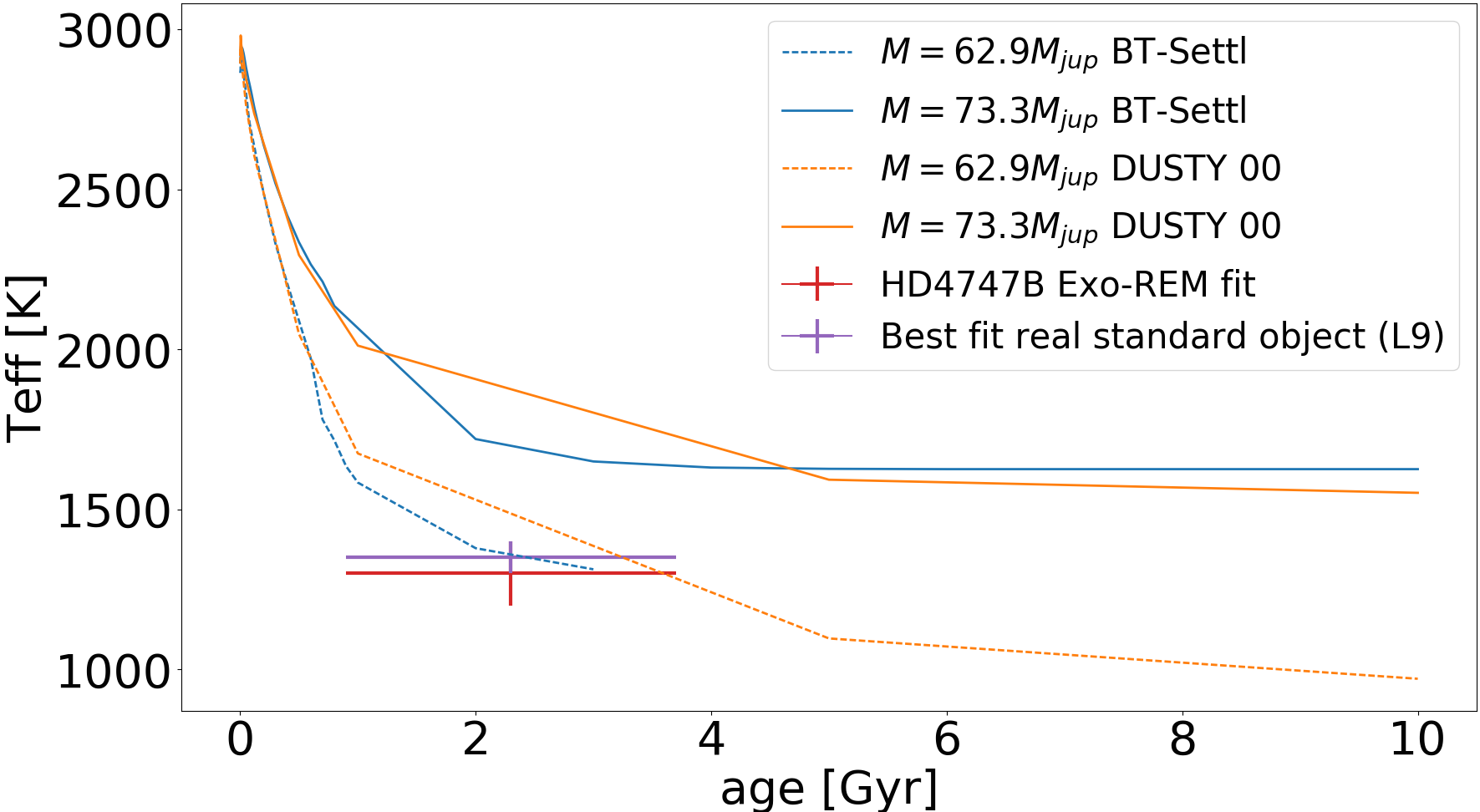}
\caption{Comparison between evolutionary models \citep[BT-Settl][]{Allard2011}, real standard objects and the Exo-REM forward modeling. The predictions from the DUSTY model \citep{Chabrier2000b} are also shown. The evolutionary models seem to underestimate the cooling rate of the sub-stellar objects.}\label{Teff_vs_age}
\end{figure}

We observe that the peak at 1.2-1.3 $\mu$m is not well fitted by the BT-Settl models, despite the good match found when comparing with the spectra of standard objects. It could mean that the opacity at this wavelength is not perfectly computed in these models, or that the grid of spectra we used is not dense enough in metallicity as well as in log($g$).

We can use our spectrum, the estimated age of the system and the model-independent dynamical mass measurement to compare with the predictions of evolutionary models. Using the measured effective temperature, which combines the information of the overall spectrum, the BHAC15 models \citep{Baraffe2015} predict a mass around 65 $\rm M_{Jup}$ (Fig. \ref{Teff_vs_age}). The BHAC15 evolutionary models tend apparently to overestimate the temperature for a given age and mass in this range of mass. In other words the object seems to cool faster than the models predict. However, the age uncertainty is quite high, and a more accurate value as well as thinner model grids at the H-burning limit would be needed to constrain correctly the models.

\section{Conclusion}\label{conclusion}
HD 4747B is a useful mass-age-metallicity benchmark object for comparison with brown dwarf atmospheric and evolutionary models. We used the SPHERE instrument to obtain high precision astrometric and spectroscopic measurements to refine its measured parameters and allow for a more thorough comparison with models.

The HD 4747AB system has been analyzed by combining radial velocity measurements and high contrast imaging. With a spectroscopic analysis we derive an effective temperature $T_{\rm eff}=5400\pm60$ K, a $\log(g)=4.60\pm0.15$ and a metallicity [Fe/H]$=-0.23\pm0.05$ dex for the primary star. A primary mass of $M_{\rm star}=0.856\pm0.014$ $\rm M_\odot$ and an age of $2.3\pm1.4$ Gyr have been also derived. Combining the SPHERE data with new radial velocity measurements from the CORALIE spectrograph, and a detection in an archival NACO dataset with previously published epochs, we derived a dynamical mass of $m_{\rm B}=70.2\pm1.6$ $\rm M_{Jup}$. 

We adjusted the spectrum extracted from the SPHERE IFS and IRDIS data, with known standard objects and derived an L9 spectral type, which is in good agreement with our color-magnitude diagram derived from the IRDIS K1 and K2 filters and previous observations. A forward analysis was conducted by using Exo-REM and confirmed an effective temperature of $T_{\rm eff}=1300\pm100$ K and a cloudy atmosphere. A radius $R=0.91\pm0.16$ $\rm R_{Jup}$ has been derived, however the $\log(g)=4\pm0.5$ found is not reproducing the dynamical mass derived in this study. HD 4747B is the most massive object analyzed with Exo-REM, and more work is needed to investigate why the gravity is apparently underestimated with this model.

A retrieval analysis allowed to derive the chemistry of the atmosphere, the Temperature-Pressure profile and the carbon and oxygen abundances (C/H=$-4.72_{-0.39}^{+0.47}$, O/H=$-3.79_{-0.08}^{+0.16}$). We compared these values to the HR 8799's planets and show that a formation scenario for HD 4747B by gravitational instability is compatible, which is favored as well by the mass ratio between the primary and its companion. 

Finally a comparison with the BT-Settl evolutionary models was conducted. The effective temperature and log($g$) derived are in good agreement with the spectral analysis in this paper. By comparing the age and the effective temperature, we obtain a model dependent mass estimation around 65 $\rm M_{Jup}$, which slightly differs to our dynamical one.

\begin{acknowledgements}
This work has been carried out within the frame of the National Center for Competence in Research "PlanetS" supported by the Swiss National Science Foundation (SNSF).\\
SPHERE is an instrument desigend and built by a consortium consisting of IPAG (Grenoble, France), MPIA (Heidelberg, Germany), LAM (Marseille, France), LESIA (Paris, France), Laboratoire Lagrange (Nice, France), INAF - Osservatorio di Padova (Italy), Observatoire Astronomique de l'Universit{\'e} de Gen{\`e}ve (Switzerland), ETH Zurich (Switzerland), NOVA (Netherlands), ON- ERA (France) and ASTRON (Netherlands) in collaboration with ESO. SPHERE was funded by ESO, with additional contributions from CNRS (France), MPIA (Germany), INAF (Italy), FINES (Switzerland) and NOVA (Netherlands). SPHERE also received funding from the European Commission Sixth and Seventh Framework Programmes as part of the Optical Infrared Coordination Network for Astronomy (OPTICON) under grant number RII3-Ct-2004-001566 for FP6 (2004-2008), grant number 226604 for FP7 (2009-2012) and grant number 312430 for FP7 (2013-2016).\\
This publication makes use of The Data \& Analysis Center for Exoplanets (DACE), which is a facility based at the University of Geneva (CH) dedicated to extrasolar planets data visualisation, exchange and analysis. DACE is a platform of the Swiss National Centre of Competence in Research (NCCR) PlanetS, federating the Swiss expertise in Exoplanet research. The DACE platform is available at https://dace.unige.ch." \\
This publication makes use of data products from the Two Micron All Sky Survey, which is a joint project of the University of Massachusetts and the Infrared Processing and Analysis Center/California Institute of Technology, funded by the National Aeronautics and Space Administration and the National Science Foundation.\\
This research has benefitted from the SpeX Prism Spectral Libraries, maintained by Adam Burgasser at http://pono.ucsd.edu/~adam/browndwarfs/spexprism.\\
This publication makes use of VOSA, developed under the Spanish Virtual Observatory project supported from the Spanish MICINN through grant AyA2011-24052.\\
J.L.B. acknowledges the support of the UK Science and Technology Facilities Council\\
Finally we acknowledge support from the "Progetti Premiali" funding scheme of the Italian Ministry of Education, University, and Research.
\end{acknowledgements}

\bibliographystyle{aa}
\bibliography{biblio}

\begin{appendix}
\section{Archival NACO observation: $\chi^2$ maps over double roll subtraction angles}\label{naco_appendix}
\begin{figure*}[!htb]
\centering
	\subfloat[$\chi^2$ map of the image with a double roll subtraction angle of $5^\circ$]{\includegraphics[width=5.8cm]{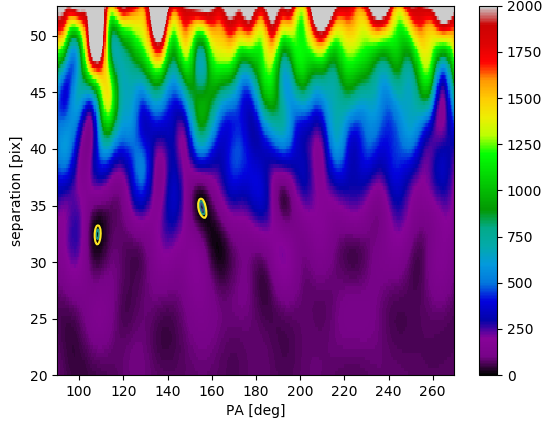}\label{chi2_5}}
\hspace{0.2 cm}
	\subfloat[$\chi^2$ map of the image with a double roll subtraction angle of $10^\circ$]{\includegraphics[width=5.8cm]{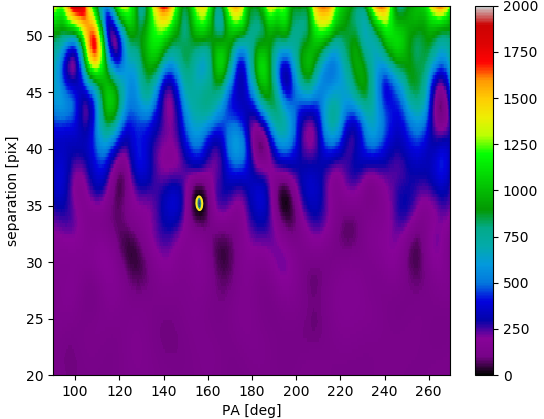}\label{chi2_10}}
\hspace{0.2 cm}
	\subfloat[$\chi^2$ map of the image with a double roll subtraction angle of $15^\circ$]{\includegraphics[width=5.8cm]{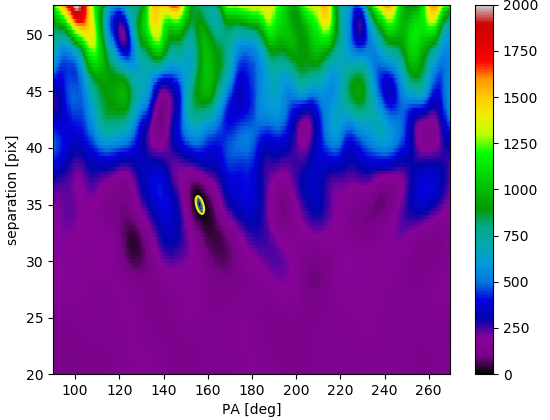}\label{chi2_15}}
\hspace{0.2 cm}
	\subfloat[$\chi^2$ map of the image with a double roll subtraction angle of $20^\circ$]{\includegraphics[width=5.8cm]{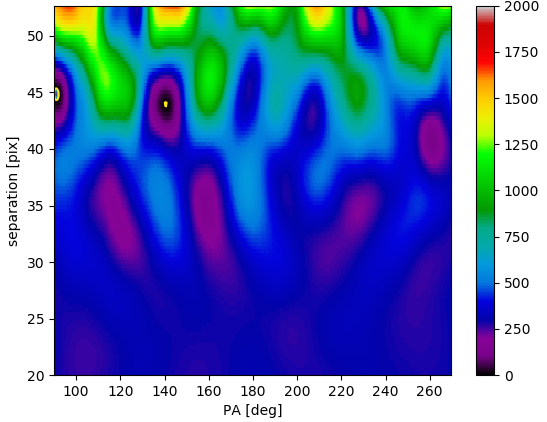}\label{chi2_20}}
\hspace{0.2 cm}
	\subfloat[$\chi^2$ map of the image with a double roll subtraction angle of $25^\circ$]{\includegraphics[width=5.8cm]{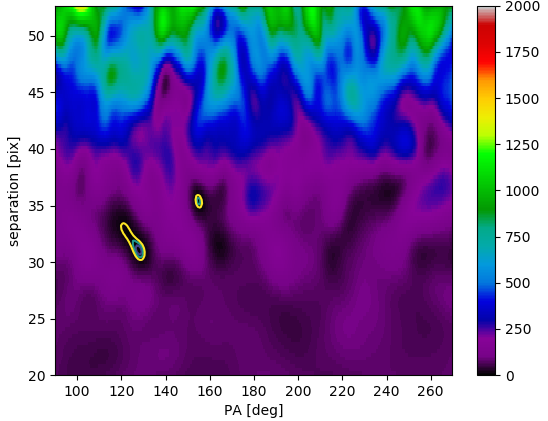}\label{chi2_25}}
\hspace{0.2 cm}
	\subfloat[Median $\chi^2$ map over double roll subtraction angles. The minimum shows the position of the companion]{\includegraphics[width=5.8cm]{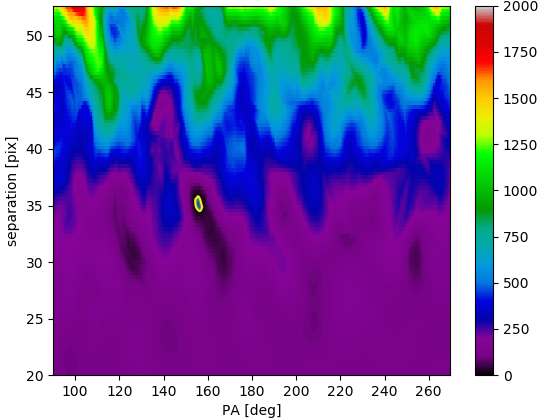}\label{chi2}}
\caption{Archival NACO data reduction. $\chi^2$ map of the image for each double roll subtraction angle. The maps shown have been computed for position angles between $90^\circ$ and $270^\circ$ and separation between 20 and 50 pixels as the companion is completely excluded in the rest of the image by the orbital analysis from the RV's, Keck and SPHERE observations.}\label{chi2_all}
\end{figure*}

As described in Sect. \ref{NACO_reduc}, we computed the $\chi^2$ maps for each double roll subtraction angle from $5^\circ$ to $25^\circ$. Each map was computed from 20 to 55 pixel of separation (346''-953'') and from $90^\circ$ to $270^\circ$, the rest of the image being excluded by the orbital information from the epochs from \citet{Crepp2016}, our SPHERE observation and the radial velocities. In Fig. \ref{chi2_all} we show these maps and the median over the double roll subtraction angles (Fig. \ref{chi2}). The position of the companion is retrieved as the minimum $\chi^2$ on each maps except for an angle of $20^\circ$ where another minimum is detected. However a local minimum is clearly identified at the companion position. This comes from the fact that the larger the angle, the least signal is on the companion. Indeed by construction, a larger double roll subtraction angle means less images to add together and thus less signal for the detection. 

\section{MCMC orbital parameters} \label{orbital_mcmc_appendix}
The MCMC simulation was performed by using $emcee$ \citep{Foreman2013}, a python stable implementation of the affine-invariant ensemble sampler for MCMC proposed by \citet{Goodman2010}. The data are modeled with a Keplerian and 4 RV offsets (one for HIRES, and 3 for the different versions of CORALIE: C98, C07, C14). The noise in the radial velocity data is modeled with a nuisance parameter for each instrument. The parallax and the mass of the primary star are also parameters of the MCMC. We ran the MCMC simulation with 39 walkers and $10^6$ steps for each walker. We computed then the correlation time scale $\tau$ of each walker and got rid of the initialization bias by removing the first $20\tau$ for each one of them \citep{Sokal1997}. To build a statistical meaningful sample, we then sampled each walker by its coherence timescale.

At the end we have 358000 independent data points which means that we characterize the parameters at a $1/\sqrt{358000}=0.17\%$ accuracy which corresponds to a $3\sigma$ confidence interval.
 
\begin{figure}[h!]
\centering
\includegraphics[width=0.97\columnwidth]{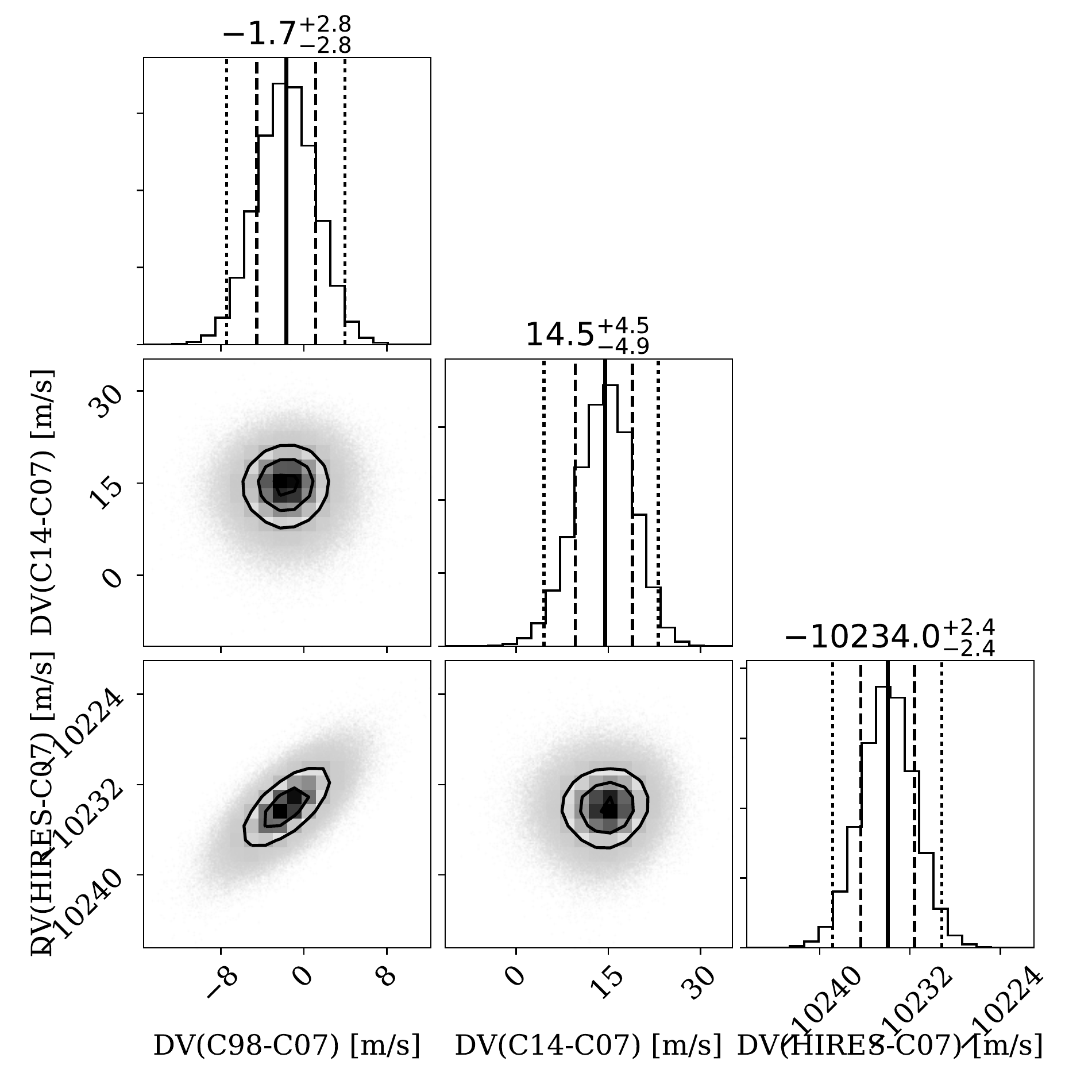}
\caption{Marginalized 1-D and 2-D posterior distributions of the orbital global adjustment combining radial velocities and direct imaging data. Here are the offsets between the different radial velocity instruments. C98 stands for the first version of the CORALIE instrument, C07 for its first update in 2007 and C14 for the last one in 2014.}\label{triangle_DV}
\end{figure}

\begin{figure*}
\centering
\includegraphics[width=1.94\columnwidth]{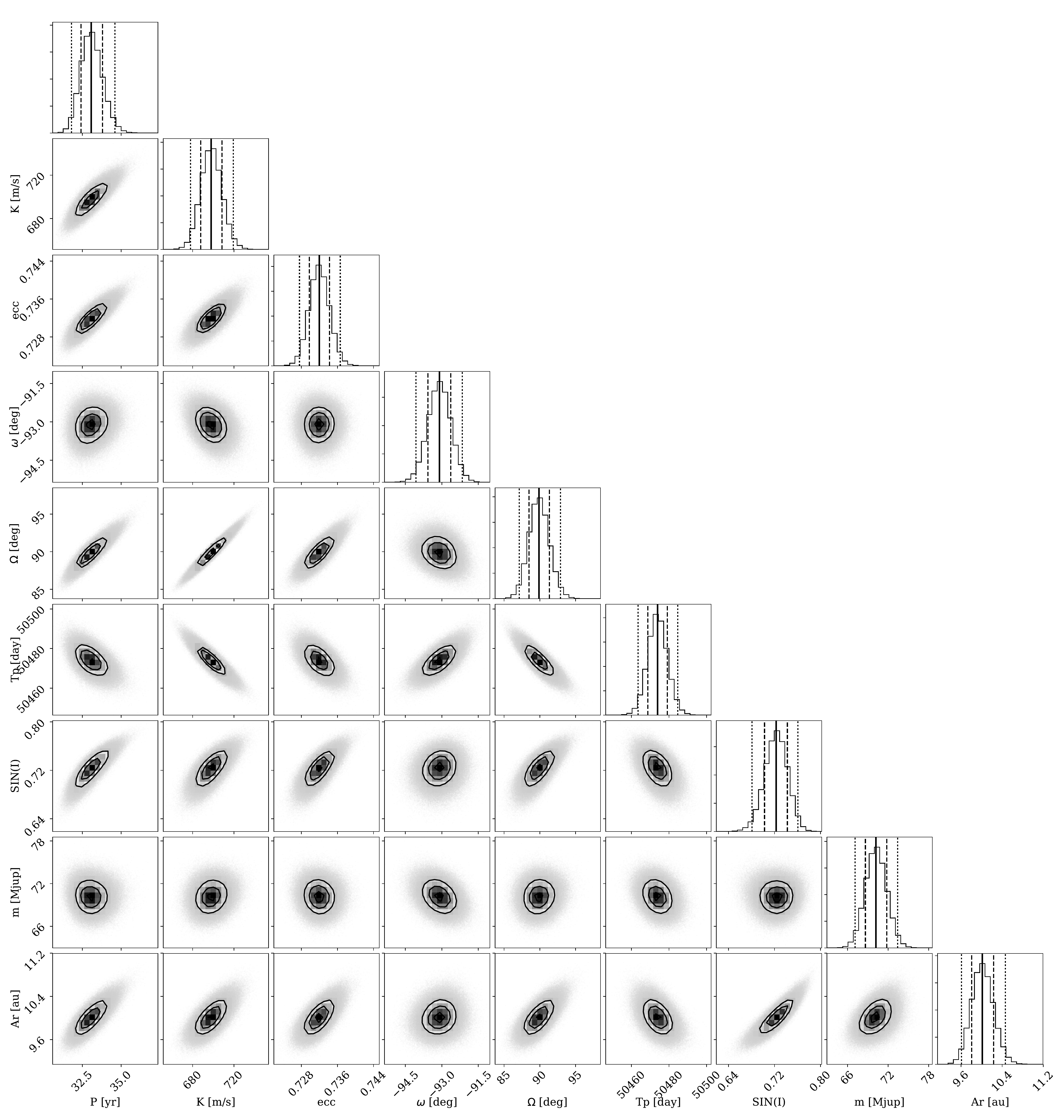}
\caption{Marginalized 1-D and 2-D posterior distributions of the orbital global adjustment combining radial velocities and direct imaging data. The model parameters P, K, e, $\omega$, $\Omega$, $T_p$ and sin(i) are shown as well as the derived quantities, the mass (m) and the orbital semi major axis (Ar).}\label{triangle_orbit}
\end{figure*}

\begin{figure*}
\centering
\includegraphics[width=1.94\columnwidth]{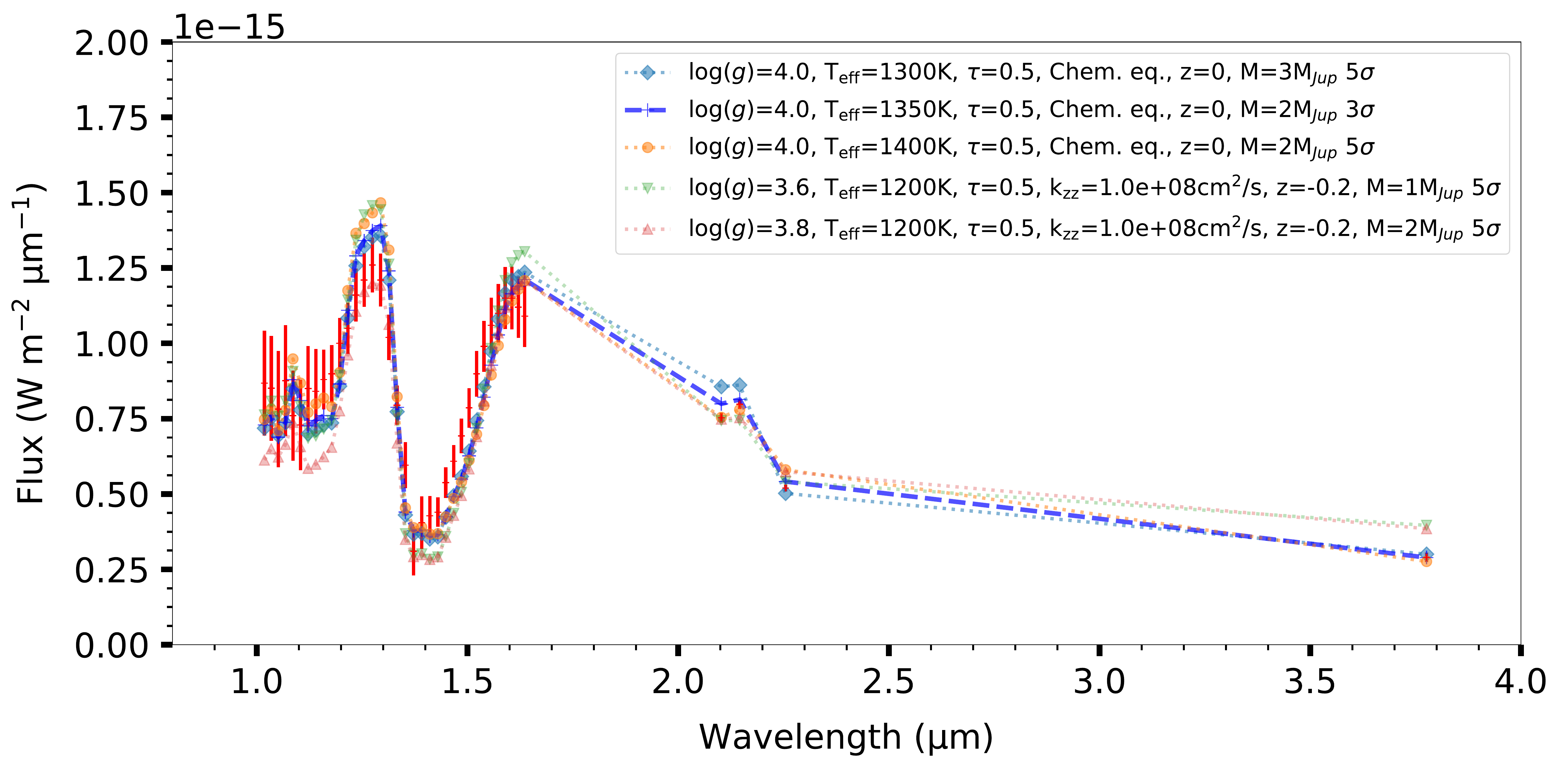}
\caption{Exo-Rem best fit to spectrum. The data are represented in red and the best fit is in light blue dashed line}\label{Exo_Rem_best_fit}
\end{figure*}

\section{Retrieval additional results}

\begin{figure*}
\centering
\includegraphics[width=2.\columnwidth]{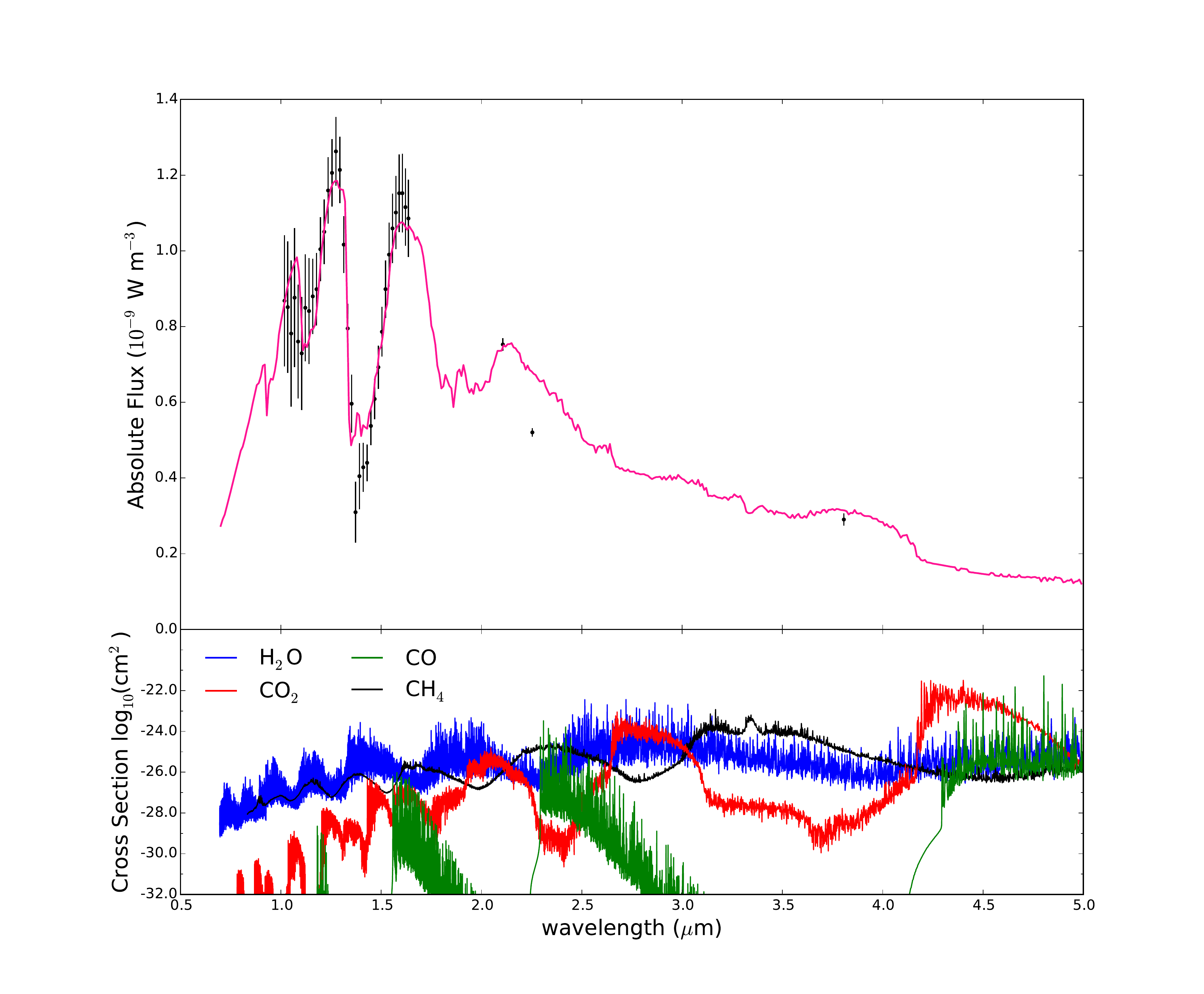}
\caption{Best-fit spectra obtained with HELIOS-R (top panel) and cross-section of the four main absorbers (bottom panel) at a pressure of 0.1 bar and a temperature of 1900 K. Lack of observations in the K-band block the detection of CO (CO absorption band between 1.5 and 2 $\mu$m )  }\label{fig:spectra}
\end{figure*}

\begin{figure*}
\centering
\includegraphics[width=.97\columnwidth]{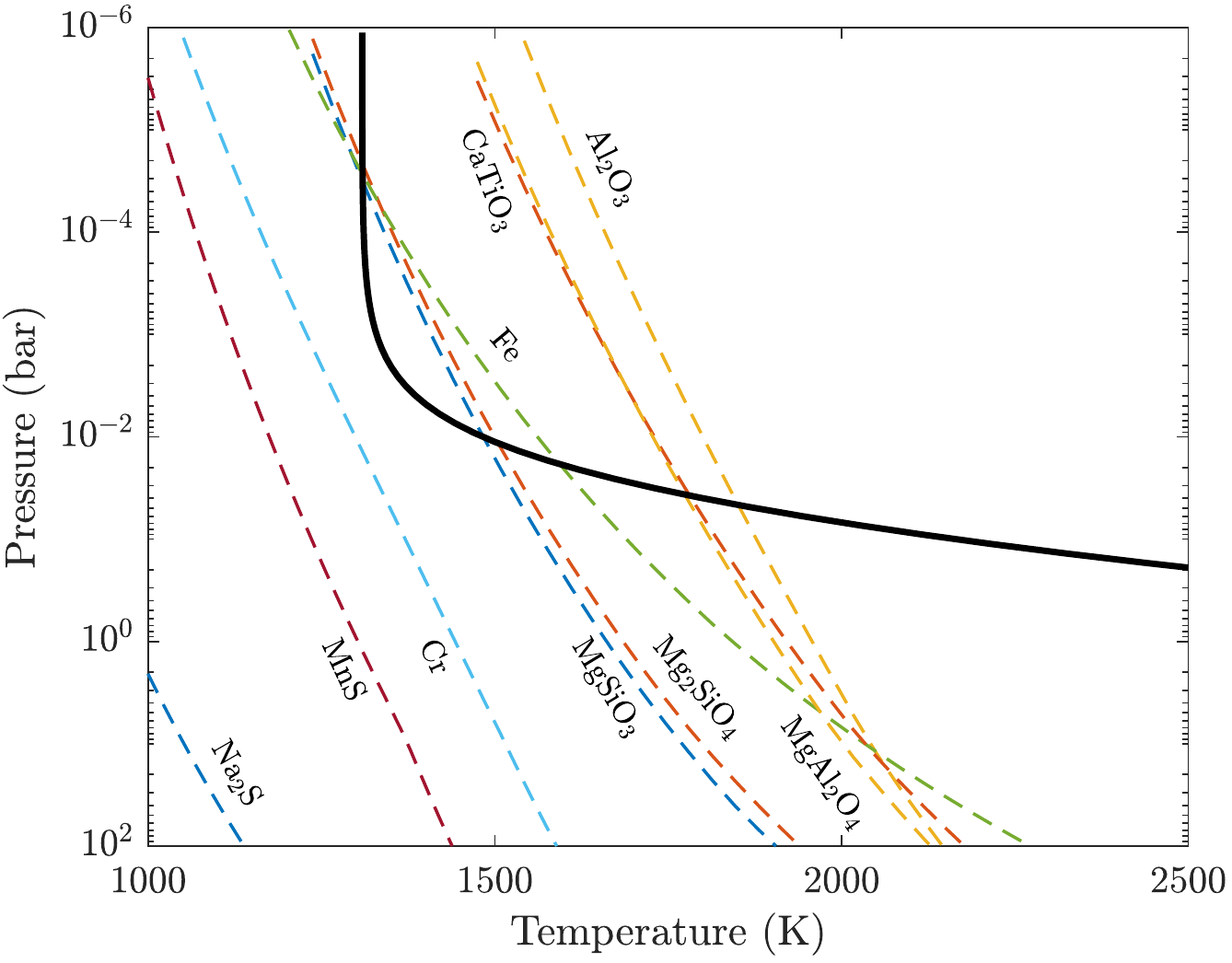}
\caption{Temperature-pressure profiles for HD 4747B obtained with HELIOS-R. Dashed lines are condensation curves for some possible condensates (assuming solar composition) }\label{fig:spectra}
\end{figure*}

\begin{figure*}
\begin{center}
\includegraphics[width=1.94\columnwidth]{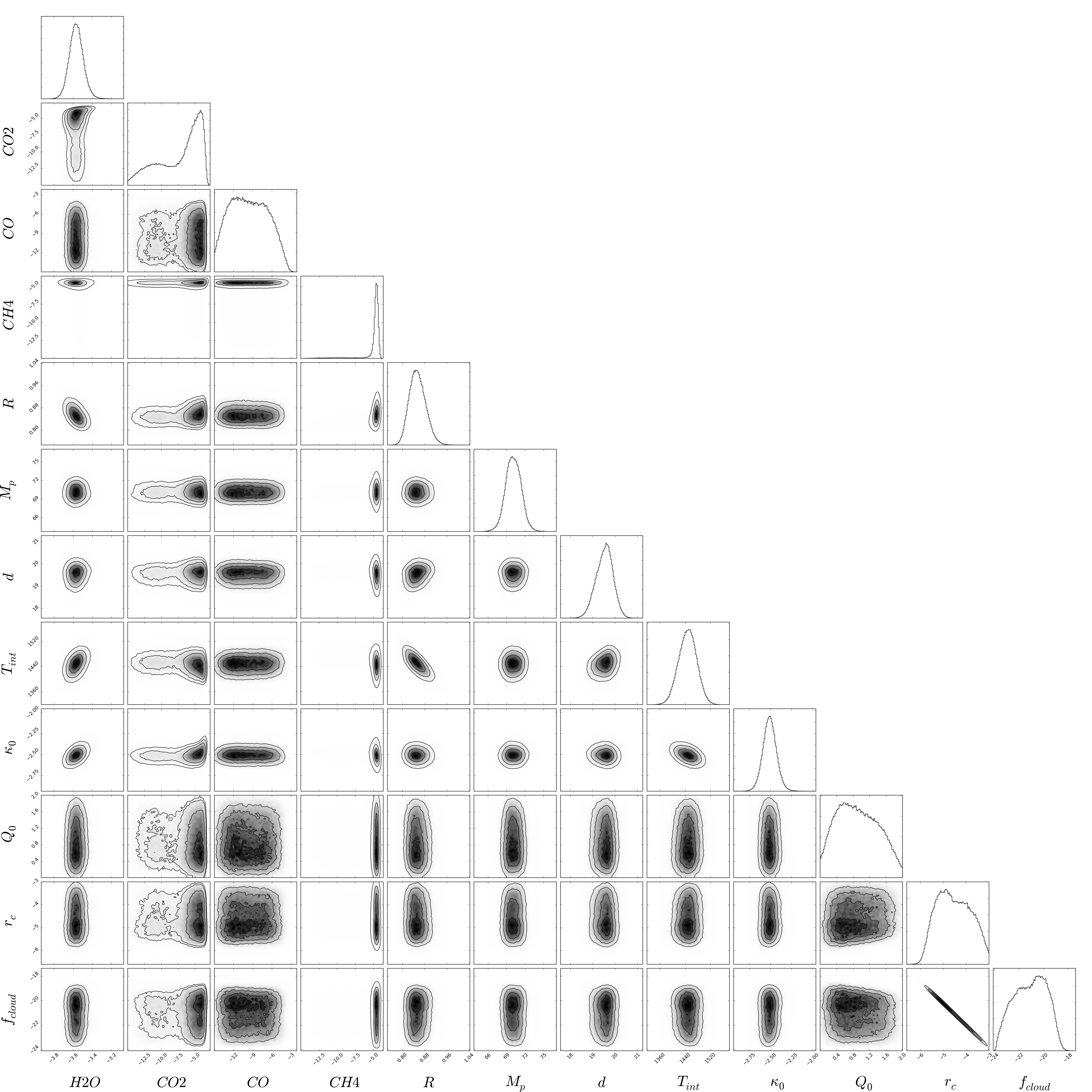}
\end{center}
\caption{Results of the Retrieval. Montage of posterior distributions from the best model (Unconstrained chemistry with clouds) of HD 4747B}
\label{fig:retrieval1}
\end{figure*}

\begin{figure}
\centering
\includegraphics[width=0.97\columnwidth]{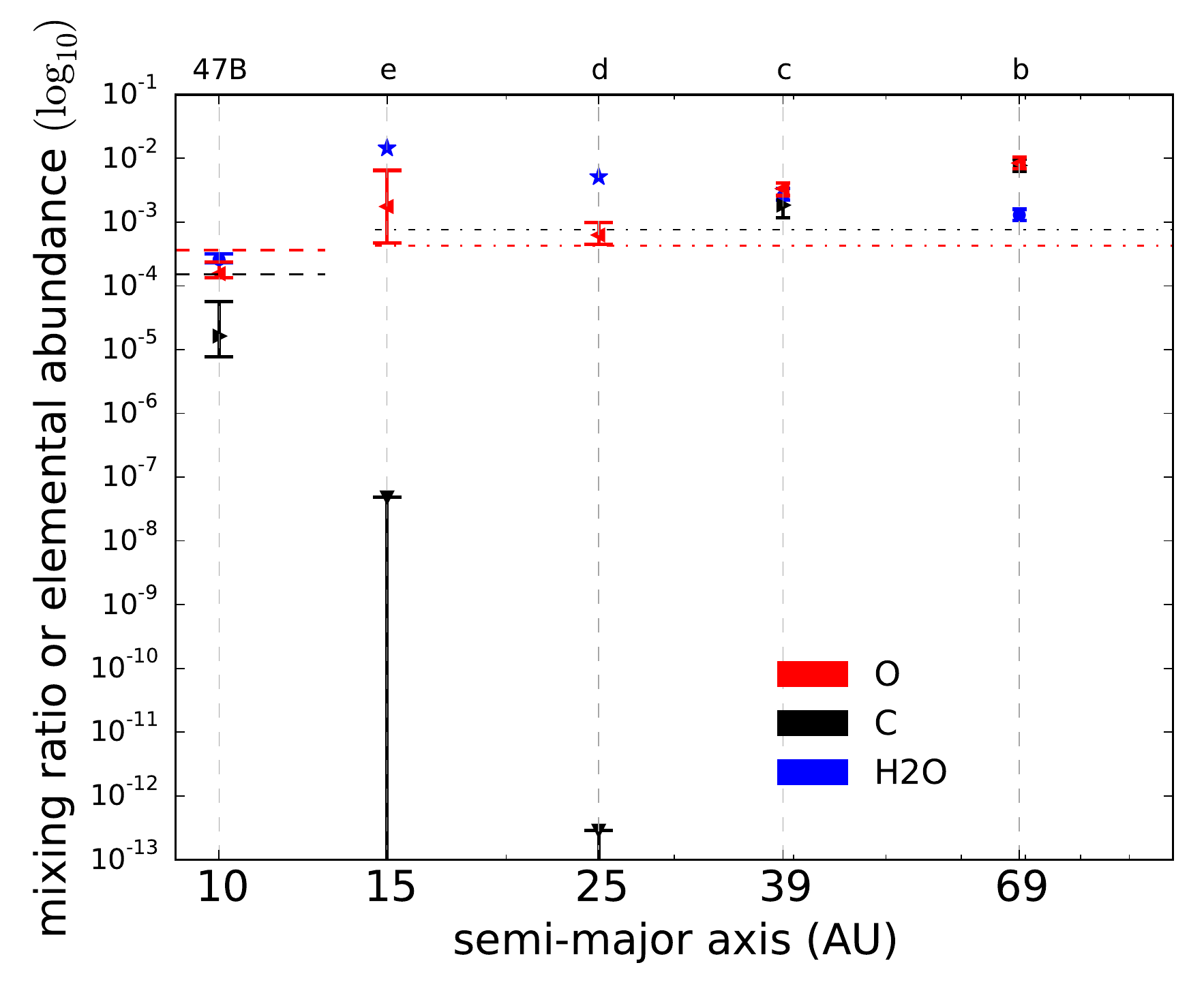}
\includegraphics[width=0.97\columnwidth]{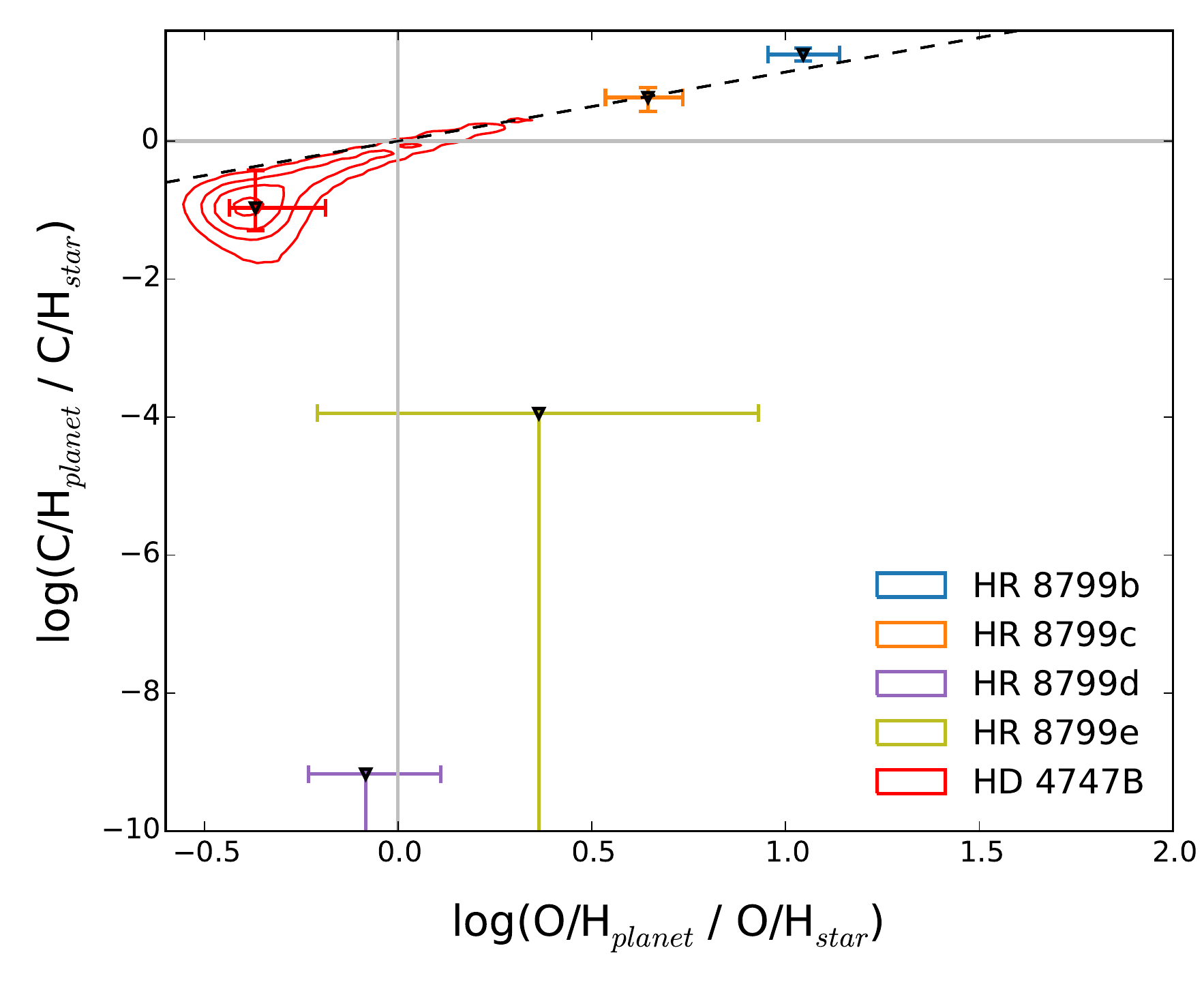}
\caption{Summary of our main results. Same as Fig. \ref{fig:summary} but with HR 8799e and HR 8799d.  The top panel shows the retrieved water mixing ratios and elemental abundances of carbon and oxygen for HD 4747B as well as the four HR 8799 exoplanets as a function of the distance to the host star . For HR 8799d and e, we show the water abundance in chemical equilibrium at 1 bar (represented by the blue stars). The carbon and oxygen abundances of the stars are shown with the dashed lines (HD 4747) and dashed dot lines(HR 8799). The bottom panel shows the companion elemental abundances normalized to its stellar values with the dashed black line denoting parity.}\label{fig:summary_2}
\end{figure}

\end{appendix}

\end{document}